\documentclass[sigconf,screen]{acmart}
\usepackage{alltt                                    
          , multirow
          , booktabs
          , listings
          , graphicx
          ,float
          ,verbatim
         ,mathtools
	   ,url
	   ,amsmath
}
\usepackage{natbib}     
\usepackage{syntax}
\usepackage{algorithm}
\usepackage{enumitem}
\usepackage{threeparttable}
\usepackage{pbox}
\usepackage[font=small,skip=1pt]{caption}
\usepackage{algpseudocode}
\usepackage{framed}
\usepackage{hhline}
\usepackage{algpseudocode}
\usepackage{balance}
\usepackage{color, colortbl}
\usepackage{tikz}
\usepackage{verbatim}
\usepackage{pifont}
\usepackage{framed}
\usetikzlibrary{arrows,shapes,backgrounds}

\usepackage{etoolbox}
\usepackage{tikz}
\usetikzlibrary{tikzmark}
\usetikzlibrary{calc}
\usepackage[all]{nowidow}

\usepackage{fancyvrb}

\tikzstyle{vertex}=[ellipse,fill=black!25,minimum size=20pt, inner sep=0pt]
\tikzstyle{edge} = [draw,thin,-]
\tikzstyle{glabel} = [text width=1cm,text centered,font=\bf]
\pgfdeclarelayer{bg}    
\pgfsetlayers{bg,main}  

\hypersetup{draft}

\usepackage{expl3}
\ExplSyntaxOn
\newcommand\latinabbrev[1]{
  \peek_meaning:NTF . {
    #1\@}%
  { \peek_catcode:NTF a {
      #1., \@ }%
    {#1., \@}}}
\ExplSyntaxOff

\algnewcommand\algorithmicswitch{\textbf{switch}}
\algnewcommand\algorithmiccase{\textbf{case}}
\algnewcommand\algorithmicassert{\texttt{assert}}
\algnewcommand\Assert[1]{\State \algorithmicassert(#1)}%
\algdef{SE}[SWITCH]{Switch}{EndSwitch}[1]{\algorithmicswitch\ #1\ \algorithmicdo}{\algorithmicend\ \algorithmicswitch}%
\algdef{SE}[CASE]{Case}{EndCase}[1]{\algorithmiccase\ #1}{\algorithmicend\ \algorithmiccase}%
\algtext*{EndSwitch}%
\algtext*{EndCase}%

\let\footnotesize\scriptsize

\algnewcommand{\LineComment}[1]{\State \(\triangleright\) #1}

\definecolor{LightGray}{rgb}{.9 .9 .9}

\newsavebox{\supbox}
\newcommand{\bsup}{\begin{lrbox}{\supbox}$\tt\scriptstyle}
\newcommand{\esup}{$\end{lrbox}{}^{\usebox{\supbox}}}
\def\eg{\latinabbrev{e.g}}
\def\ie{\latinabbrev{i.e}}

\definecolor{lightpurple}{rgb}{0.8,0.8,1}
\definecolor{codebg}{RGB}{255,255,255}
\definecolor{commentcolor}{RGB}{11,140,11}
\errorcontextlines\maxdimen

\newcommand{\ALGtikzmarkcolor}{black}
\newcommand{\ALGtikzmarkextraindent}{4pt}
\newcommand{\ALGtikzmarkverticaloffsetstart}{-.5ex}
\newcommand{\ALGtikzmarkverticaloffsetend}{-.5ex}
\makeatletter
\newcounter{ALG@tikzmark@tempcnta}

\newcommand\ALG@tikzmark@start{%
	\global\let\ALG@tikzmark@last\ALG@tikzmark@starttext%
	\expandafter\edef\csname ALG@tikzmark@\theALG@nested\endcsname{\theALG@tikzmark@tempcnta}%
	\tikzmark{ALG@tikzmark@start@\csname ALG@tikzmark@\theALG@nested\endcsname}%
	\addtocounter{ALG@tikzmark@tempcnta}{1}%
}

\def\ALG@tikzmark@starttext{start}
\newcommand\ALG@tikzmark@end{%
	\ifx\ALG@tikzmark@last\ALG@tikzmark@starttext
	\else
	\tikzmark{ALG@tikzmark@end@\csname ALG@tikzmark@\theALG@nested\endcsname}%
	\tikz[overlay,remember picture] \draw[\ALGtikzmarkcolor] let \p{S}=($(pic cs:ALG@tikzmark@start@\csname ALG@tikzmark@\theALG@nested\endcsname)+(\ALGtikzmarkextraindent,\ALGtikzmarkverticaloffsetstart)$), \p{E}=($(pic cs:ALG@tikzmark@end@\csname ALG@tikzmark@\theALG@nested\endcsname)+(\ALGtikzmarkextraindent,\ALGtikzmarkverticaloffsetend)$) in (\x{S},\y{S})--(\x{S},\y{E});%
	\fi
	\gdef\ALG@tikzmark@last{end}%
}

\apptocmd{\ALG@beginblock}{\ALG@tikzmark@start}{}{\errmessage{failed to patch}}
\pretocmd{\ALG@endblock}{\ALG@tikzmark@end}{}{\errmessage{failed to patch}}
\makeatother

\begin{document}
%

\title[Improving Bug Localization with Context-Aware Query Reformulation]{Improving IR-Based Bug Localization with Context-Aware Query Reformulation}

\author{Mohammad Masudur Rahman}
\affiliation{
\institution{University of Saskatchewan}
\country{Saskatoon, Canada}
}
\email{masud.rahman@usask.ca} 

\author{Chanchal K. Roy}
\affiliation{
\institution{University of Saskatchewan}
\country{Saskatoon, Canada}
}
\email{chanchal.roy@usask.ca}

\begin{abstract}
Recent findings suggest that Information Retrieval (IR)-based bug localization techniques do not perform well if the bug report lacks rich structured information (\eg\ relevant program entity names).
Conversely, excessive structured information (\eg\ stack traces) in the bug report might not always help the automated localization either.
In this paper, we propose a novel technique--BLIZZARD-- that automatically localizes buggy entities from project source using appropriate query reformulation and effective information retrieval.
In particular, our technique determines whether there are excessive program entities or not in a bug report (query),  
and then applies appropriate reformulations to the query for bug localization.  
Experiments using 5,139 bug reports show that our technique can localize the buggy source documents with 
7\%--56\% higher Hit@10, 6\%--62\% higher MAP@10 and 6\%--62\% higher MRR@10 than the baseline technique. Comparison with the state-of-the-art techniques and their variants report that our technique can improve 19\% in MAP@10 and 20\% in MRR@10 over the state-of-the-art, and can improve  59\% of the noisy queries and 39\% of the poor queries.   
\end{abstract}


\ccsdesc[500]{Software and its engineering~Software verification and validation}
\ccsdesc[500]{Software and its engineering~Software testing and debugging}



\keywords{Debugging automation, bug localization, bug report quality, query reformulation, information retrieval, graph-based term weighting}

\copyrightyear{2018} 
\acmYear{2018} 
\setcopyright{acmcopyright}
\acmConference[ESEC/FSE '18]{Proceedings of the 26th ACM Joint European Software Engineering Conference and Symposium on the Foundations of Software Engineering}{November 4--9, 2018}{Lake Buena Vista, FL, USA}
\acmBooktitle{Proceedings of the 26th ACM Joint European Software Engineering Conference and Symposium on the Foundations of Software Engineering (ESEC/FSE '18), November 4--9, 2018, Lake Buena Vista, FL, USA}
\acmPrice{15.00}
\acmDOI{10.1145/3236024.3236065}
\acmISBN{978-1-4503-5573-5/18/11}


\maketitle

\section{Introduction}
Despite numerous attempts for automation \cite{crowddebug,debugadvisor,majumdar,reuse,zimmer}, software debugging is still largely a manual process which costs a significant amount of development time and efforts \cite{whofix,parnin,harmful}.
One of the three steps of debugging is the identification of the location of a bug in the source code, \ie\ bug localization \cite{parnin,parninireval}. 
Recent bug localization techniques can be classified into two broad families--\emph{spectra based} and \emph{information retrieval (IR) based} \cite{locombined}.
While spectra-based techniques rely on execution traces of a software system, IR-based techniques analyse shared vocabulary between a bug report (\ie\ query) and the project source for bug localization \cite{buglocator,stacktrace}.
Performances of IR-based techniques are reported to be as good as that of spectra-based techniques, and
such performances are achieved using a low cost text analysis \cite{raobug,parninireval}. 
Unfortunately, recent qualitative and empirical studies \cite{parninireval,icse2018} have reported two major limitations.
First, IR-based techniques cannot perform well without the presence of rich structured information (\eg\ program entity names pointing to defects) in the bug reports. 
Second, they also might not perform well with a bug report that contains excessive structured information (\eg\ stack traces, Table \ref{table:noisy}) \cite{parninireval}. 
One possible explanation of these limitations could be that most of the contemporary IR-based techniques \cite{buglocator,bluir,kak,versionhistory,locombined,raobug,ldabug} use almost verbatim texts from a bug report as a \emph{query} for bug localization. That is, they do not perform any meaningful modification to the query except a limited natural language pre-processing (\eg\ stop word removal, token splitting, stemming). 
As a result, their query could be either \emph{noisy} due to excessive structured information (\eg\ stack traces) or \emph{poor} due to the lack of relevant structured information (\eg\ Table \ref{table:poor}).
One way to overcome the above challenges is to (a) refine the noisy query (\eg\ Table \ref{table:noisy}) using appropriate filters and (b) complement the poor query (\eg\ Table \ref{table:poor}) with relevant search terms. Existing studies \cite{versionhistory,wembedding,hyloc,versionhistoryjsep} that attempt to complement basic IR-based localization with costly data mining or machine learning alternatives can also equally benefit from such query reformulations.

In this paper, we propose a novel technique --BLIZZARD-- that locates software bugs from source code by employing \emph{context-aware query reformulation} and information retrieval. Our technique (1) first determines the quality (\ie\ prevalence of structured entities or lack thereof) of a bug report (\ie\ query) and classifies it as either \emph{noisy}, \emph{rich} or \emph{poor}, (2) then applies appropriate  reformulation to the query, and (3) finally uses the improved query for the bug localization with information retrieval.
Unlike earlier approaches \cite{buglocator,bluir,bluirplus,versionhistory}, it either refines a noisy query or complements a poor query for effective information retrieval.
Thus, BLIZZARD has a high potential for improving IR-based bug localization. 


To illustrate the capability of our technique in improving bug localization,
we provide two examples in which it outperforms the baseline.
The baseline technique that uses all terms except punctuation marks, stop words and digits from a bug report, returns its first correct result for the noisy query containing stack traces in Table \ref{table:noisy} at the 53$^{rd}$ position. 
On the contrary, our technique refines the same noisy query, and returns the first correct result at the first position of the ranked list which is a significant improvement over the baseline. Similarly, when we use a poor query containing no structured entities such as in Table \ref{table:poor}, the baseline technique returns the correct result at the 30$^{th}$ position. On the other hand, our technique improves the same poor query, and returns the result again at the first position.   
BugLocator \cite{buglocator}, one of the well cited IR-based techniques, returns such results at   
the 19$^{th}$ and 26$^{th}$ positions respectively for the noisy and poor queries which are far from ideal.
 
\begin{table}[!t]
	\centering
	\caption{A Noisy Bug Report (\#31637, eclipse.jdt.debug)}
	\label{table:noisy}
	\resizebox{3.4in}{!}{%
		\begin{threeparttable}
			\begin{tabular}{l|p{8cm}}
				\hline
				\textbf{Field} & \textbf{Content}\\
				\hline
				\hline
				Title & should be able to cast ``\textbf{null}"\\
				\hline
				Description & When trying to debug an application the variables \\ 
				&
				tab is empty. Also when I try to inspect or display a variable,\\ 
				& I get following error logged in the eclipse
				log file:\\
				&\begin{small}\texttt{java.lang.\textbf{NullPointerException}}\end{small}\\
				&
				\begin{small}\texttt{at org.eclipse.jdt.internal.debug.core.}\end{small}\\
				&
				\begin{small}\texttt{model.\textbf{JDIValue}.\textbf{toString}(JDIValue.java:362)}\end{small}\\
				&
				\begin{small}\texttt{at org.eclipse.jdt.internal.debug.eval.ast.}\end{small}\\
				&
				\begin{small}instructions.\textbf{Cast}.\textbf{execute}(Cast.java:88)\end{small} \\
				&
				\begin{small}\texttt{at org.eclipse.jdt.internal.debug.eval.ast.engine.}\end{small}\\
				&
				\begin{small}........................................ (10 more).......................................\end{small} \\
				\hline
			\end{tabular}
		\end{threeparttable}

	}
	\vspace{-.5cm}
\end{table}

We evaluate our technique in several different dimensions using four widely used performance metrics and 5,139 bug reports (\ie\ queries) from six Java-based subject systems. First, we evaluate in terms of the performance metrics, contrast with the baseline, and BLIZZARD localizes bugs with 7\%--56\% higher accuracy (\ie\ Hit@10), 6\%--62\% higher precision (\ie\ MAP@10) and and 6\%--62\% higher result ranks (\ie\ MRR@10) 
than the baseline (Section \ref{sec:result}). Second, we compare our technique with three bug localization techniques \cite{buglocator,bluir,versionhistoryjsep}, and our technique can improve 19\% in MAP@10 and 20\% in MRR@10 over the state-of-the-art \cite{versionhistoryjsep}  (Section \ref{sec:comparison}). Third, we also compare our approach with four state-of-the-art query reformulations techniques, and BLIZZARD improves the result ranks of 59\% of the noisy queries and 39\% of the poor queries which are 22\% and 28\% higher respectively than that of the state-of-the-art \cite{saner2017masud} (Section \ref{sec:comparison}).
By incorporating \emph{report quality aspect} and \emph{query reformulation} into IR-based bug localization, we resolve an important issue which was either not addressed properly or otherwise overlooked by earlier studies, which makes our work \emph{novel}.
Thus, the paper makes the following contributions:
\begin{itemize}[noitemsep,topsep=1pt]
\item A novel query reformulation technique that filters noise from and adds complementary information to the bug report,  
and suggests improved queries for bug localization.   
\item A novel bug localization technique that locates bugs from the project source by employing quality paradigm of bug reports, query reformulation, and information retrieval.
\item Comprehensive evaluation of the technique using 5,139 bug reports from six open source systems and validation against seven techniques including the state-of-the-art.
\item A working prototype with detailed experimental data for replication and third party reuses.
\end{itemize}


\section{Graph-Based Term Weighting}\label{sec:bg}
Term weighting is a process of determining relative importance of a term within a body of texts (\eg\ document). \citet{tfidf} first introduced TF-IDF (\ie\ term frequency $\times$ inverse document frequency) as a proxy to term importance which had been widely used by information retrieval community for the last couple of decades. 
Unfortunately, TF-IDF does not consider semantic dependencies among the terms in their importance estimation.
\citet{rada} later proposed TextRank as a proxy of term importance which was adapted from Google's PageRank \cite{pagerank} and was reported to perform better than TF-IDF.
In TextRank, a textual document is encoded into a text graph where unique words from the document are denoted as nodes, and meaningful relations among the words are denoted as connecting edges \cite{rada}. 
Such relationships could be statistical (\eg\ co-occurrence), syntactic (\eg\ grammatical modification) or semantic (\ie\ conceptual relevance) in nature \cite{blanco}.
In this research, we identify important terms using graph-based term weighting from a bug report that might contain structured elements (\eg\ stack traces) and unstructured regular texts.


\begin{table}[!t]
	\centering
	\caption{A Poor Bug Report (\#187316, eclipse.jdt.ui)}
	\label{table:poor}
	\resizebox{3.4in}{!}{%
		\begin{threeparttable}
			\begin{tabular}{l|p{8cm}}
				\hline
				\textbf{Field} & \textbf{Content}\\
				\hline
				\hline
				Title &  [preferences] Mark Occurences Pref Page\\
				\hline
				Description & There should be a link to the pref page on which you can change the color.
				Namely: General/Editors/Text Editors/Annotations. It's a pain in the a** to find the pref if you
				do not know Eclipse's preference structure well.\\
				\hline
			\end{tabular}
		\end{threeparttable}
	}
	\vspace{-.5cm}
\end{table}

\begin{figure*}
\centering
\resizebox{5.6in}{!}{%
\begin{tikzpicture}[scale=.9, auto,swap]

\node at (-1.6,2.4) [circle, inner sep=2pt, draw] (1) {\scriptsize{1}};
\node at (-.7,3.2) [circle, inner sep=1pt,draw] (2.a) {\scriptsize{2a}};
\node at (.5,2.2) [circle, inner sep=1pt, draw] (2.b) {\scriptsize{2b}};
\node at (-.6,0.2) [circle, inner sep=1pt, draw] (2.c) {\scriptsize{2c}};
\node at (1.1,3.6) [circle, inner sep=1pt,draw] (3.a) {\scriptsize{3a}};
\node at (1.9,2.1) [circle, inner sep=1pt,draw] (3.b) {\scriptsize{3b}};
\node at ((.9,0.2) [circle, inner sep=1pt,draw] (3.c) {\scriptsize{3c}};
\node at (4,3.3) [circle, inner sep=1pt,draw] (4.a) {\scriptsize{4a}};
\node at (3.9,2) [circle, inner sep=1pt,draw] (4.b) {\scriptsize{4b}};
\node at ((4,0.7) [circle, inner sep=1pt,draw] (4.c) {\scriptsize{4c}};
\node at (5.2,2.4) [circle,circle, inner sep=2pt,draw] (5) {\scriptsize{5}};
\node at (6.4,2.4) [circle,circle, inner sep=2pt,draw] (6) {\scriptsize{6}};
\node at (7.7,2.4) [circle,circle, inner sep=2pt,draw] (7) {\scriptsize{7}};
\node at (9.5,2.5) [circle,circle, inner sep=2pt,draw] (8) {\scriptsize{8}};

\begin{pgfonlayer}{bg}
\node[inner sep=0pt] (br0) at (-3.1,1.8)
    {\includegraphics[width=.25in]{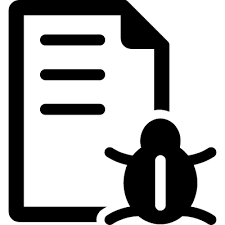}};
\node[inner sep=0pt] (br) at (-2.6,1.8)
    {\includegraphics[width=.25in]{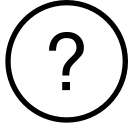}};
\node at (-3.1,1) (b) {\scriptsize Bug report};
\node at (-3.1,0.7) (b) {\scriptsize (Initial query)};

\node[inner sep=0pt] (regex) at (-1.6,1.8)
    {\includegraphics[width=.25in]{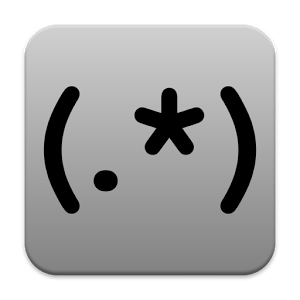}};
\node[inner sep=0pt] (cluster) at (-1,1.8)
    {\includegraphics[width=.25in]{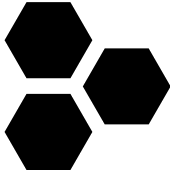}};
\node at (-1.4,1) (b) {\scriptsize Bug report};
\node at (-1.4,0.7) (b) {\scriptsize classification};
\draw[->,thick] (br) -- (regex);

\node[inner sep=0pt] (st) at (0,3.2)
    {\includegraphics[width=.30in]{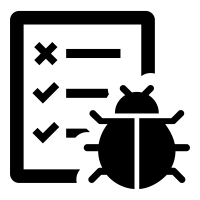}};
\node at (0,2.5) (b) {\scriptsize BR$_{ST}$};
\draw[->,thick] (cluster) -- (st);

\node[inner sep=0pt] (pe) at (0,1.8)
    {\includegraphics[width=.30in]{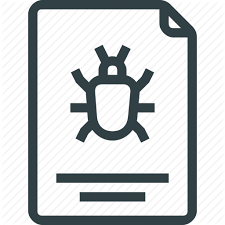}};
\node at (0,1.2) (b) {\scriptsize BR$_{PE}$};
\draw[->,thick] (cluster) -- (pe);

\node[inner sep=0pt] (nl) at (0,0.5)
    {\includegraphics[width=.30in]{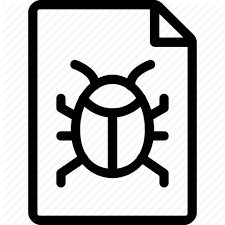}};
\node at (0,-0.1) (b) {\scriptsize BR$_{NL}$};
\draw[->,thick] (cluster) -- (nl);

\node[inner sep=0pt] (err) at (1.5,3.2)
    {\includegraphics[width=.30in]{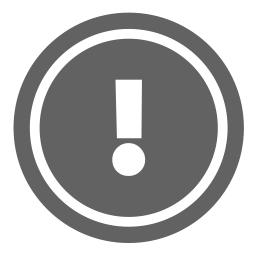}};
\node[inner sep=0pt] (trace) at (2.1,3.2)
    {\includegraphics[width=.30in]{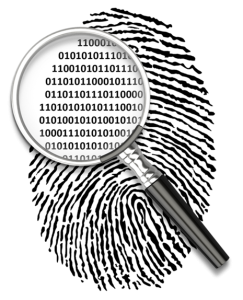}};
\node at (1.8,2.5) (b) {\scriptsize Exception \& traces};
\draw[->,thick] (st) -- (err);

\node[inner sep=0pt] (tracegraph) at (3.4,3.2)
    {\includegraphics[width=.35in]{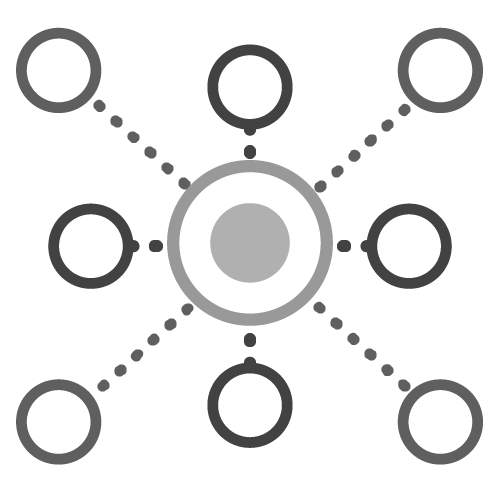}};
\node at (3.4,2.5) (b) {\scriptsize Trace graph};
\draw[->,thick] (trace) -- (tracegraph);

\node[inner sep=0pt] (prep) at (1.5,1.8)
    {\includegraphics[width=.28in]{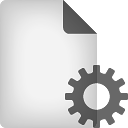}};
\node at (1.5,1.2) (b) {\scriptsize Text preprocessing};
\draw[->,thick] (pe) -- (prep);

\node[inner sep=0pt] (termgraph) at (3.4,1.8)
    {\includegraphics[width=.35in]{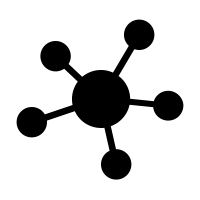}};
\node at (3.4,1.2) (b) {\scriptsize Text graph};
\draw[->,thick] (prep) -- (termgraph);

\node[inner sep=0pt] (prf) at (1.4,0.5)
    {\includegraphics[width=.30in]{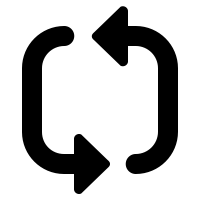}};
\node[inner sep=0pt] (code) at (2.2,0.5)
    {\includegraphics[width=.28in]{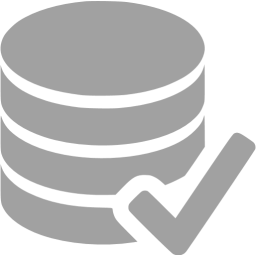}};
\node at (1.8,-0.1) (b) {\scriptsize Pseudo-relevance};
\node at (1.8,-0.4) (b) {\scriptsize feedback};
\draw[->,thick] (nl) -- (prf);

\node[inner sep=0pt] (tokengraph) at (3.4,0.5)
    {\includegraphics[width=.35in]{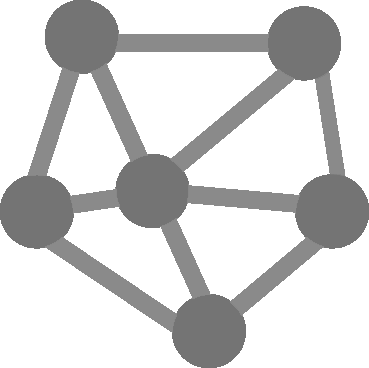}};
\node at (3.4,-0.1) (b) {\scriptsize Source token};
\node at (3.4,-0.4) (b) {\scriptsize graph};
\draw[->,thick] (code) -- (tokengraph);

\node[inner sep=0pt] (pagerank) at (5,1.8)
    {\includegraphics[width=.30in]{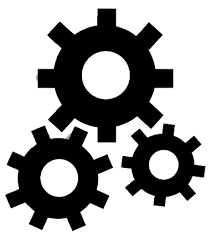}};
\node at (5,1.1) (b) {\scriptsize Graph-based};
\node at (5,0.8) (b) {\scriptsize term weighting};
\draw[->,thick] (tracegraph) -- (pagerank);
\draw[->,thick] (termgraph) -- (pagerank);
\draw[->,thick] (tokengraph) -- (pagerank);

\node[inner sep=0pt] (ranking) at (6.4,1.8)
    {\includegraphics[width=.30in]{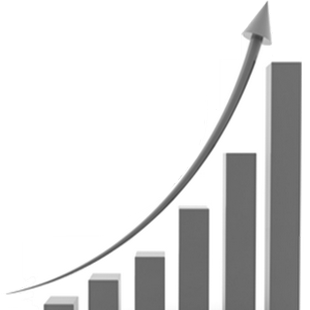}};
\node at (6.4,1.1) (b) {\scriptsize Term ranking};
\draw[->,thick] (pagerank) -- (ranking);

\node[inner sep=0pt] (reformulated) at (7.7,1.8)
    {\includegraphics[width=.25in]{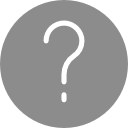}};
\node at (7.8,1.1) (b) {\scriptsize Reformulated};
\node at (7.8,0.8) (b) {\scriptsize query};
\draw[->,thick] (ranking) -- (reformulated);

\node[inner sep=0pt] (se) at (8.9,1.8)
    {\includegraphics[width=.30in]{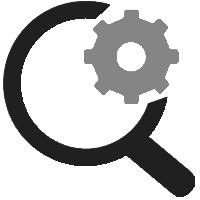}};
\node[inner sep=0pt] (cb) at (9.6,1.9)
    {\includegraphics[width=.28in]{./sysdiag/codebase.png}};
\node at (9.3,1.1) (b) {\scriptsize Bug};
\node at (9.3,0.8) (b) {\scriptsize localization};
\draw[->,thick] (reformulated) -- (se);

\node[inner sep=0pt] (bf1) at (8.9,3.2)
    {\includegraphics[width=.25in]{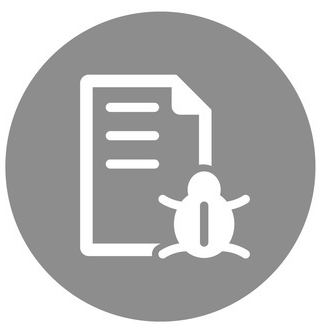}};
\node[inner sep=0pt] (bf2) at (9.6,3.2)
    {\includegraphics[width=.25in]{./sysdiag/bf1.png}};
\node at (9.1,3.7) (b) {\scriptsize Buggy entities};
\draw[->,thick] (se) -- (bf1);

\draw[dotted] (-4.1,-.6) rectangle (.5,3.9);
\draw[dotted] (.6,-.6) rectangle (8.2,3.9);
\draw[dotted] (8.3,-.6) rectangle (10.1,3.9);

\node at (-3.7,-.2) [circle, inner sep=1pt,draw] (A) {\textbf{A}};
\node at (5.7,-.2) [circle, inner sep=1pt,draw] (B) {\textbf{B}};
\node at (8.7,-.2) [circle, inner sep=1pt,draw] (C) {\textbf{C}};

\end{pgfonlayer}
\end{tikzpicture}
}
\vspace{.2cm}
\caption{Schematic diagram of the proposed technique: (A) Bug report classification, (B) Query reformulation, and (C) Bug localization}
\label{fig:sysdiag}
\vspace{-.3cm}
\end{figure*}

\section{BLIZZARD: Proposed Technique} \label{sec:blizzard}

Fig. \ref{fig:sysdiag} shows the schematic diagram of our proposed technique--BLIZZARD. Furthermore, Algorithm \ref{algo} shows the pseudo-code for BLIZZARD. We make use of bug report quality, query reformulation, and information retrieval for localizing bugs in source code from bug reports of any quality as shown in the following sections:


\subsection{Bug Report Classification}\label{sec:classification}
Since our primary objective with this work is to overcome the challenges posed by the different kinds of information bug reports may contain, we categorize the reports prior to bug localization. In addition to having natural language texts, a bug report typically may contain different structured elements: (1) stack traces (reported active stack frames during the occurrence of a bug, \eg\ Table \ref{table:noisy}), and (2) program elements such as method invocations, package names, and source file names. Having consulted with the relevant literature \cite{structcls,parninireval,goodbugreport}, we classify the bug reports into three board categories (Steps 1, 2a, 2b and 2c, Fig. \ref{fig:sysdiag}) as follows:


\textbf{BR$\mathbf{_{ST}}$:} $ST$ stands for stack traces. If a bug report contains one or more stack traces besides the regular texts or program elements, it is classified into BR$_{ST}$. Since trace entries contain too much structured information,  query generated from such a report is generally considered \emph{noisy}. 
We apply the following regular expression \cite{stacktrace} to locate the trace entries from the report content. 

\begin{small}
\begin{Verbatim*}[frame=single]
(.*)?(.+)\.(.+)(\((.+)\.java:\d+\)|\(Unknown Source\)
|\(Native Method\))
\end{Verbatim*}
\end{small}

\textbf{BR$\mathbf{_{PE}}$:} $PE$ stands for program elements. If a bug report contains one or more program elements (\eg\ method invocations, package names, source file name) but no stack traces in the texts, then it is classified into BR$_{PE}$. Queries generated from such report are considered \emph{rich}. We use appropriate regular expressions \cite{rigby} to identify the program elements from the texts. 


\textbf{BR$\mathbf{_{NL}}$:} $NL$ stands for natural language. If a bug report contains neither any program elements nor any stack traces, it is classified into BR$_{NL}$. That is, it contains only unstructured natural language description of the bug. Queries generated from such reports are generally considered \emph{poor} in this work.   

We adopt a semi-automated approach in classifying the bug reports (\ie\ the queries). Once a bug report is provided, we employ each of our regular expressions to determine its class. If the automated step fails due to ill-defined structures of the report, the class is determined based on manual analysis. Given the explicit nature of the structured entities, human developers can identify the class easily. The contents of each bug report are considered as the \emph{initial queries} which are reformulated in the next few steps.


\subsection{Query Reformulation}
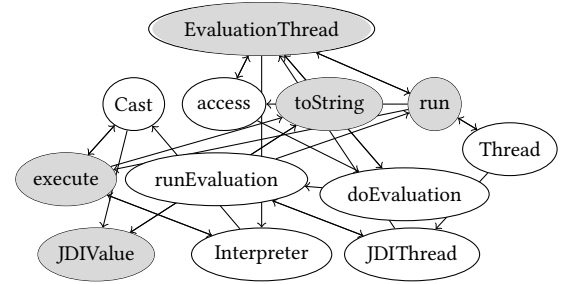
\begin{figure}
	\centering
	\begin{tikzpicture}[scale=1, auto,swap]
	\small
	\foreach \pos/\name in {
		{(-.1,0)/Cast},
		{(1.1,0)/access},
		{(1.6,-2)/Interpreter},
		{(-.6,-2)/JDIValue},
		{(2.5,0)/toString},
		{(3.9,0)/run},
		{(1,-1)/runEvaluation},
		{(3.5,-1.2)/doEvaluation},
		{(1.6,1)/EvaluationThread},
		{(-1,-1)/execute},
		{(3.6,-2)/JDIThread},
		{(4.9,-.6)/Thread}}
	\node[vertex] (\name) at \pos [draw=black, inner sep=1pt, fill=white] {\small\name};
	
	\node[vertex] at (1.6,1) [fill=gray!30] {\small EvaluationThread};
	\node[vertex] at (2.5,0) [inner sep=1pt,fill=gray!30] {\small toString};
	\node[vertex] at (-.6,-2) [inner sep=1pt,fill=gray!30] {\small JDIValue};
	\node[vertex] at (3.9,0) [inner sep=1pt,fill=gray!30] {\small run};
	\node[vertex] at (-1,-1) [inner sep=1pt,fill=gray!30] {\small execute};
	
	\begin{pgfonlayer}{bg}    
	\foreach \source/ \dest  in {
		execute/Interpreter,
		JDIValue/toString,
		run/Thread,
		JDIThread/EvaluationThread,
		Cast/execute,
		runEvaluation/run,
		EvaluationThread/Interpreter,
		run/access,
		execute/toString,
		Thread/JDIThread,
		Cast/JDIValue,
		runEvaluation/JDIThread,
		run/execute,
		EvaluationThread/access,
		execute/Cast,
		EvaluationThread/doEvaluation,
		access/doEvaluation,
		Interpreter/execute,
		EvaluationThread/run,
		doEvaluation/EvaluationThread,
		toString/JDIValue,
		Thread/run,
		Interpreter/Cast,
		access/EvaluationThread,
		doEvaluation/runEvaluation,
		run/EvaluationThread,
		JDIThread/runEvaluation}
	\path[edge, ->, draw=black] (\source) -- (\dest);
	\end{pgfonlayer}
	
	\end{tikzpicture}
	\vspace{.1cm}
	\caption{Trace graph of stack traces in Table \ref{table:noisy}}
	\label{fig:tgraph}
	\vspace{-.7cm}
\end{figure}

Once bug reports (\ie\ queries) are classified into three classes above  based on their structured elements or lack thereof, we apply appropriate reformulations to them. In particular, we analyse either bug report contents or the results retrieved by them, employ graph-based term weighting, and then identify important keywords from them for query reformulation as follows: 


\textbf{Trace Graph Development from BR$\mathbf{_{ST}}$:}\label{sec:tracegraph}
According to existing findings \cite{icse2018,parninireval}, bug reports containing stack traces are potentially noisy, and performances of the bug localization using such reports (\ie\ the queries) are below the average. Hence, important \emph{search keywords} should be \emph{extracted} from the noisy queries for effective bug localization. In this work, we transform the stack traces into a trace graph (\eg\ Fig. \ref{fig:tgraph}) (Steps 3a, 4a, Fig. \ref{fig:sysdiag}, Lines 8--10, Algorithm \ref{algo}), and identify the important keywords using a graph-based term weighting algorithm namely PageRank \cite{rada,blanco}.



To the best of our knowledge, to date, graph-based term weighting has been   
employed only on unstructured regular texts \cite{saner2017masud} and semi-structured source code \cite{ase2017masud}. 
On the contrary, we deal with stack traces which are structured and should be analysed carefully. 
Stack traces generally comprise of an error message containing the encountered exception(s), and an ordered list of method invocation entries.   
Each invocation entry can be considered as a tuple $t\{P,C,M\}$  that contains a package name $P$, a class name $C$, and a method name $M$.
While these entities are statically connected within a tuple, they are often hierarchically connected (\eg\ caller-callee relationships) to other tuples from the traces as well.
\citet{hillicse09} consider method signatures and field signatures as salient entities from the source code, and suggest keywords from them for code search.
Similarly, we consider class name and method name from each of the $N$ tuples as the salient items, and represent them as the nodes and their dependencies as the connecting edges in the graph. 
In stack traces, the topmost entry (\ie\ $i=1$) has the highest degree of interest \cite{context} which gradually decreases for the entries at the lower positions in the list.
That is, if $t_{i}\{P_i,C_i,M_i\}$ is a tuple under analysis, and $t_{j}\{P_j,C_j,M_j\}$ is a neighbouring tuple with greater degree of interest, then the nodes $V_i$ and edges $E_i$ are added to the trace graph $G_{ST}$ as follows: 
\begin{equation*}
\small
\setlength{\abovedisplayskip}{1pt}
\setlength{\belowdisplayskip}{1pt}
\begin{split}
V_i=\{C_i, M_i\},~E_i= \{C_i\leftrightarrow M_i\}\cup\{C_i\rightarrow C_j, M_i\rightarrow M_j\}\mid j=i-1\\
V=\bigcup\limits_{i=1}^ N\{V_{i}\},~E=\bigcup\limits_{i=1}^N\{E_{i}\},~G_{ST}=(V,E) 
\end{split}
\end{equation*}
For the example stack traces in Table \ref{table:noisy}, the following connecting edges: 
\texttt{JDIValue}$\leftrightarrow$\texttt{toString}, \texttt{Cast}$\leftrightarrow$\texttt{execute}, \texttt{Cast}$\rightarrow$\texttt{JDIValue}, \\\texttt{execute}$\rightarrow$\texttt{toString}, \texttt{Interpreter}$\leftrightarrow$\texttt{execute}, and \texttt{Interpreter}\\$\rightarrow$\texttt{Cast}  
 are added to the example trace graph in Fig. \ref{fig:tgraph}.

\textbf{Text Graph Development from BR$\mathbf{_{PE}}$:}
Bug reports containing relevant program entities (\eg\ method names) are found effective as queries for IR-based bug localization \cite{parninireval,bluir,icse2018}.
However, we believe that \emph{appropriate keyword selection} from such reports can further \emph{boost up} the localization performance.
Existing studies employ TextRank and POSRank on natural language texts, and identify search keywords for concept location \cite{saner2017masud} and information retrieval \cite{rada,blanco}. 
Although bug reports (\ie\ from BR$_{PE}$) might contain certain structures such as program entity names (\eg\ class name, method name) and code snippets besides natural language texts, the existing techniques could still be applied to them given that these structures are treated appropriately. 
We thus remove stop words \cite{stopword} and programming keywords \cite{keyword} from a bug report, \emph{split} the structured tokens using \emph{Samurai} (\ie\ a state-of-the-art token splitting tool \cite{samurai}), and then transform the preprocessed report ($R_{pp}$) into a set of sentences ($S\in R_{pp}$).
We adopt \citet{saner2017masud} that exploits \emph{co-occurrences} and \emph{syntactic dependencies} among the terms for identifying important terms from a textual body (\eg\ change request).
We thus develop two text graphs (Steps 3b, 4b, Fig. \ref{fig:sysdiag}, Lines 10--11, Algorithm \ref{algo}) using co-occurrences and syntactic dependencies among the words from each report as follows:

\emph{(1) Text Graph using Word Co-occurrences:} In natural language texts, 
the semantics (\ie\ senses) of a given word are often determined by its contexts (\ie\ surrounding words) \cite{wordsense, wembedding,w2vec}.
That is, co-occurring words complement the semantics of each other. 
We thus consider a sliding window of size $K$ (\eg\ $K=2$) \cite{rada}, 
capture co-occurring words, and then encode the word co-occurrences within each window into connecting edges $E$ of a text graph \cite{saner2017masud}. The individual words ($\forall w_i\in V$) are denoted as nodes in the graph. Thus, for a target word $w_i$, the following node $V_i$  and two edges $E_i$ will be added to the text graph G$_{PE}$ as follows:
\begin{equation*}
\setlength{\abovedisplayskip}{1pt}
\setlength{\belowdisplayskip}{1pt}
\begin{split}
V_i= \{w_i\},~E_i=\{w_i\leftrightarrow w_{i-1}, w_i\leftrightarrow w_{i+1}\}\mid S=[w_1..w_i..w_N] \\
V=\bigcup\limits_{\forall S\in R_{pp}} \bigcup\limits_{w_i\in S}\{V_{i}\},~E=\bigcup\limits_{\forall S\in R_{pp}}\bigcup\limits_{w_i\in S}\{E_{i}\},~G_{PE}=(V,E)
\end{split}
\end{equation*}
Thus, the example phrase--\emph{``source code directory"}--yields two edges, \emph{``source"}$\leftrightarrow$\emph{``code"} and \emph{``code"}$\leftrightarrow$\emph{``directory"} 
while extending the text graph with three distinct nodes-- \emph{``source", ``code"} and \emph{``directory"}.  

\emph{(2) Text Graph using POS Dependencies:} According to \emph{Jespersen's Rank} theory \cite{jespersen,blanco,saner2017masud}, 
parts of speech (POS) from a sentence can be divided into three ranks-- \emph{primary} (\ie\ noun), \emph{secondary} (\ie\ verb, adjective) and \emph{tertiary} (\ie\ adverb)-- where words from a higher rank generally define (\ie\ modify) the words from the same or lower ranks. That is, a noun modifies only another noun whereas a verb modifies another noun, verb or an adjective.
We determine POS tags using Stanford POS tagger \cite{postagger}, and    
encode such syntactic dependencies among words into connecting edges and individual words as nodes in a text graph.
For example, the sentence annotated using Penn Treebank tags \cite{postagger}--\emph{``Open$_{VB}$ the$_{DT}$ source$_{NN}$ code$_{NN}$ directory$_{NN}"$}--has the following syntactic dependencies: \emph{``source"}$\leftrightarrow$\emph{``code"}, \emph{``code"}$\leftrightarrow$\emph{``directory"}, \emph{``source"}$\leftrightarrow$\emph{``directory"}, \\\emph{``open"}$\leftarrow$\emph{``source"},
\emph{``open"}$\leftarrow$\emph{``code"}   and \emph{``open"}$\leftarrow$\emph{``directory"}, and thus adds six connecting edges to the text graph. 
 
\textbf{Source Term Graph Development for BR$\mathbf{_{NL}}$:}
Bug reports containing only natural language texts and no structured entities are found not effective for IR-based bug localization \cite{parninireval,icse2018}. We believe that such bug reports possibly miss the right keywords for bug localization.
Hence, they need to be \emph{complemented} with \emph{appropriate keywords} before using. A recent study \cite{ase2017masud} provides improved reformulations to a poor natural language query for concept location by first collecting \emph{pseudo-relevance feedback} and then employing graph-based term weighting. In pseudo-relevance feedback, 
Top-K result documents, returned by a given query, are naively considered as relevant and hence, are selected for query reformulation \cite{qsurvey,refoqus}.
Since bug reports from BR$_{NL}$ class contain only natural language texts, the above study might directly be applicable to them.
We thus adopt their approach for our query reformulation, collect Top-K (\eg\ $K=10$) source code documents retrieved by a BR$_{NL}$-based query, and develop a source term graph (Steps 3c, 4c, Fig. \ref{fig:sysdiag}, Lines 13--15, Algorithm \ref{algo}).  

\citet{hillicse09} consider method signatures and fields signatures from source code as the salient items, and suggest keywords for code search from them.
In the same vein, we also collect these signatures from each of the $K$ feedback documents for query reformulation. In particular, we extract structured tokens from each signature, split them using \emph{Samurai}, and then generate a natural language phrase from each token \cite{hillicse09}. For example, the method signature--\texttt{getContextClassLoader()}--can be represented as a verbal phrase-- \emph{``get Context Class Loader"}.
We then analyse such phrases across all the feedback documents, capture co-occurrences of terms within a fixed window (\ie\ $K=2$) from each phrase, and develop a source term graph. Thus, the above phrase adds four distinct nodes and three connecting edges -- \emph{``get"}$\leftrightarrow$\emph{``context"}, \emph{``context"}$\leftrightarrow$\emph{``class"} and \emph{``class"}$\leftrightarrow$\emph{``loader"} -- to the source term graph.
\begin{algorithm}[!tb]
\caption{Bug Localization with Query Reformulation and IR}
\label{algo}
\small
\begin{algorithmic}[1]
\Procedure{BLIZZARD} {$R$}\Comment{$R$: a given bug report}
\State $Q' \gets$ \{\}\Comment{reformulated query terms}
\LineComment{Classifying and preprocessing the bug report $R$}
\State $C_R \gets$ getBugReportClass($R$)
\State $R_{pp}\gets$ preprocess ($R$)
\LineComment{Representing the bug report as a graph}
\Switch{$C_R$}
    \Case{BR$_{ST}$}
      \State $ST \gets$ getStackTraces ($R$)
      \State $G_{ST} \gets$ getTraceGraph ($ST$) 
    \EndCase
    \Case{BR$_{PE}$}
      \State $G_{PE} \gets$ getTextGraphs ($R_{pp}$)
    \EndCase
    \Case{BR$_{NL}$}
	\State $R_F \gets$ getPseudoRelevanceFeedback ($R_{pp}$)
      \State $G_{NL} \gets$ getSourceTermGraph ($R_F$)
    \EndCase
\EndSwitch
\LineComment{Getting term weights and search keywords}
\If{ClassKey $CK \in$ \{$ST,PE,NL$\} } 
\State $PR_{CK}\gets$ getPageRank ($G_{CK}$)
\State $Q[C_R] \gets$ getTopKTerm(sortByWeight($PR_{CK}$))
\EndIf
\LineComment{Constructing the reformulated query $Q'$}
\Switch{$C_R$}
\Case{BR$_{ST}$}
\State $N_E \gets$ getExceptionName($R$)
\State $M_E \gets$ getErrorMessage($R$)
\State $Q' \gets$ \{$N_E \cup  M_E \cup Q[C_R]$\}
 \EndCase
\Case{BR$_{PE}$}
\State $Q' \gets Q[C_R]$
 \EndCase
\Case{BR$_{NL}$}
\State $Q' \gets$ \{$R_{pp} \cup Q[C_R]$\}
 \EndCase
\EndSwitch
\LineComment{Bug localization with $Q'$ from codebase $corpus$}
\State \textbf{return} Lucene($corpus$, $Q'$)
\EndProcedure
\end{algorithmic}
\end{algorithm}
\setlength{\textfloatsep}{2pt}

\textbf{Term Weighting using PageRank:} Once each body of texts (\eg\ stack traces, regular texts, source document) is transformed into a graph, we apply PageRank \cite{pagerank, rada,saner2017masud,ase2017masud} to the graph for identifying important keywords. PageRank was originally designed for web link analysis, and it determines the reputation of a web page based on the votes or recommendations (\ie\ hyperlinks) from other reputed pages on the web \cite{pagerank}.
Similarly, in the context of our developed graphs, the algorithm determines importance of a node (\ie\ term) based on incoming links from other important nodes of the graph. In particular, it analyses the connectivity (\ie\ connected neighbours and their weights) of each term $V_i$ in the graph recursively, and then calculates the node's weight $TW(V_i)$: 
\begin{equation*}\label{eq:textrank}
\small
\setlength\abovedisplayskip{1pt}
\setlength\belowdisplayskip{1pt}
TW(V_{i})=(1-\phi)+\phi\sum_{j\epsilon In(V_{i})}\frac{TW(V_{j})}{|Out(V_{j})|}~~ (0 \le \phi \le1)
\end{equation*}
Here, $In(V_{i})$ refers to nodes providing incoming links to $V_i$, $Out(V_{j})$ refers to nodes that $V_{j}$ is connected to through outgoing links, and $\phi$ is the damping factor. 
\citet{pagerank} consider $\phi$ as the probability of staying on the web page and $1-\phi$ as the probability of jumping off the page by a random surfer. 
They use $\phi=0.85$ which was adopted by later studies \cite{rada,blanco,saner2017masud}, and we also do the same.      
We initialize each node in the graph with a value of 0.25 \cite{rada}, and recursively calculate their weights unless they converge below a certain threshold (\ie\ 0.0001) or the iteration count reaches the maximum  (\ie\ 100) \cite{rada}.
Once the calculation is over, we end up with an accumulated weight for each node (Step 5, Fig. \ref{fig:sysdiag}, Lines 16--20, Algorithm \ref{algo}). Such weight of a node is considered as an estimation of relative importance of corresponding term among all the terms (\ie\ nodes) from the bug report (\ie\ graph).

\textbf{Reformulation of the Initial Query:} 
Once term weights are calculated, we rank the terms based on their weights, and select the Top-K (8$\le K \le$30, Fig. \ref{fig:query-length}) terms for query reformulations. 
Since bug reports (\ie\ initial queries) from three classes have different degrees of structured information (or lack thereof), we carefully apply our reformulations to them (Steps 6, 7, Fig. \ref{fig:sysdiag}, Lines 21--30, Algorithm \ref{algo}).
In case of BR$_{ST}$ (\ie\ noisy query), we replace trace entries with the reformulation terms, extract the error message(s) containing exception name(s), and combine them as the reformulated query.
For BR$_{NL}$ (\ie\ poor query), we combine preprocessed report texts with the highly weighted source code terms as the reformulated query.   
In the case of BR$_{PE}$, only Top-K weighted terms from the bug report are used as a reformulated query for bug localization.


\subsection{Bug Localization}
\textbf{Code Search:} Once a reformulated query is constructed, we submit the query to \emph{Lucene} \cite{refoqus,trconfig}. Lucene is a widely adopted search engine for document  search that combines Boolean search and VSM-based search methodologies (\eg\ TF-IDF \cite{tfidf}). In particular, we employ the Okapi BM25 similarity from the engine,
use the reformulated query for the code search, and then collect the results (Step 8, Fig. \ref{fig:sysdiag}, Lines 31--32, Algorithm \ref{algo}). These resultant and potentially buggy source code documents are then presented as a ranked list to the developer for manual analysis.

\textbf{Working Examples:} Table \ref{table:we} shows our reformulated queries for the showcase bug reports in Table \ref{table:noisy} (\ie\ BR$_{ST}$), Table \ref{table:poor} (\ie\ BR$_{NL}$), and another example report from BR$_{PE}$ class. Baseline queries from these reports return their first correct results at the 53$^{rd}$ (for BR$_{ST}$), 27$^{th}$ (for BR$_{PE}$) and 30$^{th}$ (for BR$_{NL}$) positions of their corresponding ranked lists. On the contrary, BLIZZARD refines the noisy query from BR$_{ST}$ report, selects important keywords from BR$_{PE}$ report, and enriches the poor query from BR$_{NL}$ report by adding complementary terms from relevant source code. 
As a result, all three reformulated queries return their first correct results (\ie\ buggy source files) at the topmost (\ie\ first) positions, which demonstrate the potential of our technique for bug localization.

\begin{table}[!t]
\centering
\caption{Working Examples}\label{table:we}
\resizebox{3.2in}{!}{%
\begin{threeparttable}
\begin{tabular}{l|l|p{5.6cm}|c}
\hline
\textbf{Technique} & \textbf{Group} & \textbf{Query Terms} & \textbf{QE}\\
\hline
\hline
Baseline & \multirow{2}{*}{BR$_{ST}$} & 127 terms from Table \ref{table:noisy} after preprocessing, \textbf{Bug ID\# 31637, eclipse.jdt.debug} & 53 \\
\hhline{-~--}
BLIZZARD &  & \texttt{NullPointerException} + \emph{``Bug should be able to cast \texttt{null}"} + \{\texttt{JDIValue toString execute EvaluationThread run}\}  & \textbf{01}\\
\hline
\hline
Baseline & \multirow{2}{*}{BR$_{PE}$} & 195 terms (after preprocessing) from \textbf{Bug ID\#  15036, eclipse.jdt.core} & 27 \\
\hhline{-~--}
BLIZZARD &  & \{astvisitor post postvisit previsit pre file post pre astnode visitor\} &  \textbf{01} \\
\hline
\hline
Baseline & \multirow{2}{*}{BR$_{NL}$} & 32 terms from Table \ref{table:poor} after preprocessing, \textbf{Bug ID\# 475855, eclipse.jdt.ui} & 30\\
\hhline{-~--}
BLIZZARD &  & Preprocessed report texts + \{\texttt{compliance create preference add configuration field dialog annotation}\} & \textbf{01}\\
\hline
\end{tabular}
\centering
\textbf{QE} = Query Effectiveness, rank of the first returned correct result
\end{threeparttable}
}
\vspace{.2cm}
\end{table}

\begin{table*}
	\centering
	\caption{Experimental Dataset}\label{table:expds}
	\resizebox{7in}{!}{%
		\begin{threeparttable}
			\begin{tabular}{l|c|c|c|c|c||l|c|c|c|c|c||l}
				\hline
				\textbf{System} & \textbf{Time Period}  & \textbf{BR$\mathbf{_{ST}}$} & \textbf{BR$\mathbf{_{PE}}$} & \textbf{BR$\mathbf{_{NL}}$} & \textbf{BR$\mathbf{_{All}}$} & \textbf{System}& \textbf{Time Period} & \textbf{BR$\mathbf{_{ST}}$} & \textbf{BR$\mathbf{_{PE}}$} & \textbf{BR$\mathbf{_{NL}}$} & \textbf{BR$\mathbf{_{All}}$} & \textbf{Total}\\
				\hline
				\hline
				ecf & Oct, 2001--Jan, 2017 & 71 & 319 & 163 & 553  & eclipse.jdt.ui & Oct, 2001--Jun, 2016 & 130 & 578 & 407 & 1,115 & BR$_{ST}$ = \textbf{826} (16.06\%)   \\
				\hline
				eclipse.jdt.core & Oct, 2001--Sep, 2016  & 159 & 698 & 132 & 989 & 
				eclipse.pde.ui & Oct, 2001--Jun, 2016 & 123 & 239 & 510 & 872 
				& BR$_{PE}$ = \textbf{2,767} (53.81\%) \\
				\hline
				eclipse.jdt.debug &Oct, 2001--Jan, 2017 & 126 & 202 & 229 & 557& tomcat70 &Sep, 2001--Aug, 2016 & 217 & 731 & 105 & 1,053
				& BR$_{NL}$= \textbf{1,546} (30.08\%) \\
				\hline
				\multicolumn{13}{c}{\textbf{Total:} 5,139}\\
				\hline
			\end{tabular}
			\centering
			\textbf{BR$_{ST}$}=Bug reports with stack traces, \textbf{BR$_{PE}$}=Bug reports with program entities but no stack traces, \textbf{BR$_{NL}$}=Bug reports with only natural language texts
		\end{threeparttable}
	}
	\vspace{-.4cm}
\end{table*}

\section{Experiment}\label{sec:experiment}

We evaluate our proposed technique in several different dimensions using four widely used performance metrics and more than 5K bug reports (the queries) from six different subject systems. First, we evaluate in terms of the performance metrics and contrast with the baseline for different classes of bug reports/queries (Section \ref{sec:result}). Second, we compare our approach with three state-of-the-art bug localization techniques (Section \ref{sec:comparison}). Third, and possibly the most importantly, we also compare our approach with four state-of-the-art query reformulations techniques (Section \ref{sec:comparison}). In particular, we answer four research questions using our experiments as follows:

  
\begin{itemize}
\item \textbf{RQ$\mathbf{_1}$:} (a) How does BLIZZARD perform in bug localization, and (b) how do various parameters affect its performance?
\item \textbf{RQ$\mathbf{_2}$:} Do our reformulated queries perform better than the baseline search queries from the bug reports?
\item \textbf{RQ$\mathbf{_3}$:} Can BLIZZARD outperform the existing bug localization techniques including the state-of-the-art?
\item \textbf{RQ$\mathbf{_4}$:} Can BLIZZARD outperform the existing query reformulation techniques targeting concept/feature location and bug localization?
\end{itemize}

\subsection{Experimental Dataset}
\label{sec:dataset}
\textbf{Dataset Collection:} We collect a total of 5,139 bug reports from six open source subject systems for our experiments.
The dataset was taken from an earlier empirical study \cite{icse2018}.
Table \ref{table:expds} shows our dataset.
First, all the resolved (\ie\ marked as RESOLVED) bug reports of each subject system were collected from the BugZilla and JIRA repositories given that they were submitted within a specific time interval (Table \ref{table:expds}). 
Then the version control history of each system at GitHub was consulted to identify the bug-fixing commits \cite{bugid}. 
Such approach was regularly adopted by the relevant literature \cite{twkraft,buglocator,stacktrace}, and we also follow the same. 
In order to ensure a fair evaluation, we also discard such bug reports from our dataset for which no source code files (\eg\ Java classes) were changed or no relevant source files exist in the collected system snapshot.

\textbf{Goldset Development:}
We collect \emph{changeset} (\ie\ list of changed files) from each of our selected bug-fixing commits, and develop a \emph{goldset}. Multiple changesets for the same bug were merged together.   

\textbf{Replication Package:} Our working prototype and experimental data are publicly available \cite{blizzard} for replication and reuse.

\subsection{Performance Metrics}\label{sec:pmetrics}
We use four performance metrics for the evaluation and comparison of our technique. Since these metrics were frequently used by the relevant literature \cite{buglocator,bluir,stacktrace,versionhistory,wembedding,saner2017masud}, they are also highly appropriate for our experiments in this work.

\textbf{Hit@K:} It is defined as the percentage of queries for which at least one buggy file (\ie\ from the goldset) is correctly returned within the Top-K results. It is also called Recall@Top-K \cite{bluir} and Top-K Accuracy \cite{saner2017masud} in the literature.

\textbf{Mean Average Precision@K (MAP@K)}: Unlike regular precision, this metric considers the ranks of correct results within a ranked list. Precision@K calculates precision at the occurrence of each buggy file in the list. Average Precision@K (AP@K) is defined as the average of Precision@K for all the buggy files in a ranked list for a given query. Thus, Mean Average Precision@K is defined as the mean of Average Precision@K (AP@K) of all queries as follows: 
\begin{equation*}\label{eq:avep}
\setlength\abovedisplayskip{1pt}
\setlength\belowdisplayskip{1pt}
AP@K=\frac{\sum_{k=1}^{D}P_{k}\times buggy(k)}{|S|},~~ MAP@K=\frac{\sum_{q\epsilon Q}AP@K(q)}{|Q|}
\end{equation*}
Here, function $buggy(k)$ determines whether $k^{th}$ file (or result) is buggy (\ie\ returns 1) or not (\ie\ returns 0), $P_{k}$ provides the precision at $k^{th}$ result, and $D$ refers to the number of total results. $S$ is the gold set for a query, and $Q$ is the set of all queries. The bigger the MAP@K value is, the better a technique is.

\textbf{Mean Reciprocal Rank@K (MRR@K)}: Reciprocal Rank@K is defined as the multiplicative inverse of the rank of first correctly returned buggy file (\ie\ from gold set) within the Top-K results. Thus, Mean Reciprocal Rank@K (MRR@K) averages such measures for all queries in the dataset as follows:
\begin{equation*}
\setlength{\abovedisplayskip}{1pt}
\setlength{\belowdisplayskip}{1pt}
MRR@K(Q) =\frac{1}{|Q|}\sum_{q\in Q}{\frac{1}{firstRank(q)}}
\end{equation*}
Here, $firstRank(q)$ provides the rank of first buggy file within a ranked list. MRR@K can take a maximum value of 1 and a minimum value of 0.  The bigger the MRR@K value is, the better a bug localization technique is.

\textbf{Effectiveness (E)}: It approximates a developer's effort in locating the first buggy file in the result list \cite{stacktrace}.
That is, the measure returns the rank of first buggy file in the result list. 
The lower the effectiveness value is, the better a given query is, \ie\ the developer needs to check less amount of results from the top before reaching the actual buggy file in the list.

\subsection{Experimental Results}\label{sec:result}
We first show the performance of our technique in terms of appropriate metrics (RQ$_1$-(a)), then discuss the impacts of different adopted parameters upon the performance (RQ$_1$-(b)), and finally show our comparison with the baseline queries (RQ$_2$) as follows: 

\begin{table}
	\centering
	\caption{Performance of BLIZZARD in Bug Localization}\label{table:blizzard}
	\resizebox{3.35in}{!}{%
		\begin{threeparttable}
			\begin{tabular}{l|l|c|c|c|c|c}
				\hline
				\textbf{Dataset} & \textbf{Technique} & \textbf{Hit@1} & \textbf{Hit@5} & \textbf{Hit@10} & \textbf{MAP@10} & \textbf{MRR@10}\\
				\hline
				\hline
				\multirow{2}{*}{BR$_{ST}$} & Baseline & 21.67\% & 40.03\% & 48.25\% & 28.09\% & 0.29 \\
				\hhline{~------}
				 & BLIZZARD &  \textbf{*34.42}\% & \textbf{*66.28}\% & \textbf{*75.21}\% & \textbf{*45.50}\% & \textbf{*0.47}\\
				\hline
				\multirow{2}{*}{BR$_{PE}$} & Baseline & 39.85\% & 64.29\% & 72.09\% & 47.28\% & 0.50 \\
				\hhline{~------}
				& BLIZZARD & \textbf{44.31}\% & \textbf{*69.48}\% & \textbf{*77.84}\% & \textbf{*52.08}\% & \textbf{*0.55}\\
				\hline
				\multirow{2}{*}{BR$_{NL}$} & Baseline & 28.24\% & 50.96\% & 61.23\% & 35.48\% & 0.38\\
				\hhline{~------}
				& BLIZZARD & \textbf{29.16}\% & \textbf{53.78}\% & \textbf{65.21}\% & \textbf{*37.62}\% & \textbf{0.40} \\
				\hline
				\hline
				\multirow{2}{*}{All} & Baseline & 34.32\% & 57.83\% & 66.47\% & 41.66\% & 0.44\\
				\hhline{~------}
				& BLIZZARD & \textbf{*38.58}\% & \textbf{*65.08}\% & \textbf{*74.52}\% & \textbf{*47.13}\% & \textbf{*0.50} \\
				\hline
			\end{tabular}
			\centering
			\textbf{*}=Significantly higher than baseline, \textbf{Emboldened}= Comparatively higher   
		\end{threeparttable}
		\vspace{-.4cm}
	}
\end{table}

\textbf{Selection of Baseline Queries, and Establishment of Baseline Technique and Baseline Performance:}
Existing studies suggest that text retrieval performances could be affected by query quality \cite{refoqus}, underlying retrieval engine \cite{trconfig} or even text preprocessing steps \cite{stemming,kevic}. Hence, we choose the baseline queries and baseline technique pragmatically for our experiments.
We conduct a detailed study where three independent variables-- 
bug report field (\eg\ title, whole texts), retrieval engine (\eg\ Lucene \cite{refoqus}, Indri \cite{bluir}) and text preprocessing step (\ie\ stemming, no stemming)--are alternated, and then we choose the best performing configuration as the baseline approach. In particular, we chose the preprocessed version (\ie\ performed stop word and punctuation removal, split complex tokens but avoided stemming) of the whole texts (\ie\ \emph{title} + \emph{description}) from a bug report as a baseline query. Lucene was selected as the baseline technique since it outperformed Indri on our dataset. The performance of Lucene with the baseline queries was selected as the baseline performance (\ie\ Table \ref{table:blizzard}) for IR-based bug localization in this study. In short, our baseline is: (preprocessed whole texts + splitting of complex tokens + Lucene search engine).  

\textbf{Answering RQ$\mathbf{_1}$(a) -- Performance of BLIZZARD:}
 As shown in Table \ref{table:blizzard}, on average, our technique--BLIZZARD--localizes 74.52\% of the bugs from a dataset of 5,139 bug reports with 47\% mean average precision@10 and a mean reciprocal rank@10 of 0.50 which are 12\%, 13\% and 14\% higher respectively than the baseline performance measures. That is, on average, our technique can return the first buggy file at the second position of the ranked list, almost half of returned files are buggy (\ie\ true positive) and it succeeds three out of four times in localizing the bugs. Furthermore, while the baseline technique is badly affected by the noisy (\ie\ BR$_{ST}$) and poor queries (\ie\ BR$_{NL}$), our technique overcomes such challenges with appropriate query reformulations, and provides significantly higher performances. For example, the baseline technique can localize 48\% of the bugs from BR$_{ST}$ dataset (\ie\ noisy queries) with only 28\% precision  when Top-10 results are considered. On the contrary, our technique localizes 75\% of the bugs with 46\% precision in the same context which are 56\% and 62\% higher respectively than the corresponding baseline measures.
 Such improvements are about 7\% for BR$_{NL}$, \ie\ poor queries.
 In the cases where bug reports contain program entities, \ie\ BR$_{PE}$, and the baseline performance measures are already pretty high, our technique further refines the query and provides even higher performances. For example, BLIZZARD improves both baseline MRR@10 and baseline MAP@10 for BR$_{PE}$ dataset by 10\% which is promising.

 \begin{figure}[!t]
 	\centering
 	\includegraphics[width=3in ]{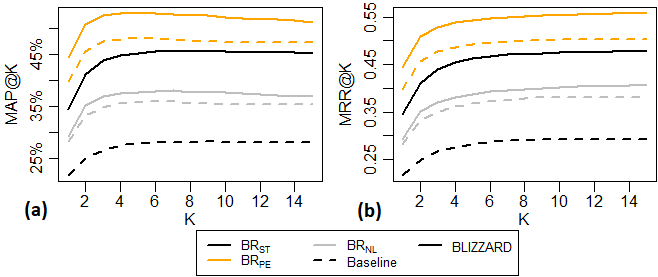}
 	\caption{Comparison of BLIZZARD with baseline technique in terms of (a) MAP@K and (b) MRR@K}
 	\label{fig:compare-baseline}
 	\vspace{-.3cm}
 \end{figure}
 
 \begin{figure}[!t]
 	\centering
 	\includegraphics[width=3in ]{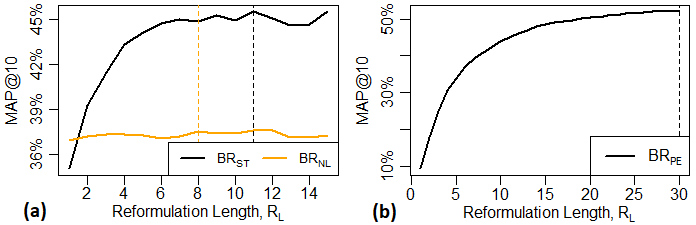}
 	\caption{Impact of query reformulation length on the MAP@10 of our technique--BLIZZARD}
 	\label{fig:query-length}
 	\vspace{-.3cm}
 \end{figure}
      
      \begin{table}
      	\centering
      	\caption{Query Improvement by BLIZZARD over Baseline Queries}\label{table:baseline-qe}
      	\resizebox{3.35in}{!}{%
      		\begin{threeparttable}
      			\begin{tabular}{l|l|c|c|c}
      				\hline
      				\textbf{Dataset} & \textbf{Query Pair} & \textbf{Improved/MRD} & \textbf{Worsened/MRD} & \textbf{Preserved}\\
      				\hline
      				\multirow{2}{*}{BR$_{ST}$} & \textbf{BLIZZARD} vs. BL$_T$ & 484 (\textbf{58.60}\%)/-82 & 206 (24.94\%)/+34 & 136 (16.46\%) \\
      				\hhline{~----}
      				& \textbf{BLIZZARD} vs. BL & 485 (\textbf{58.72}\%)/-122 & 174 (21.07\%)/+72 & 167 (20.22\%)\\ 
      				\hline
      				\multirow{2}{*}{BR$_{PE}$} & \textbf{BLIZZARD} vs. BL$_T$ & \textbf{1,397} (\textbf{50.49}\%)/-60 & 600 (21.68\%)/+38 & 770 (27.83\%) \\
      				\hhline{~----}
      				& BLIZZARD vs. BL & 865 (31.26\%)/-34 & 616 (22.26\%)/+24 & 1,286 (46.48\%)\\
      				\hline
      				\multirow{2}{*}{BR$_{NL}$} & \textbf{BLIZZARD} vs. BL$_T$ &
      				869 (\textbf{56.21}\%)/-27 & 355 (22.96\%)/+29 & 322 (20.83\%) \\
      				\hhline{~----}
      				& BLIZZARD vs. BL & 597 (38.62\%)/-16 & 455 (29.43\%)/+31 &	494 (31.95\%) \\
      				\hline
      				\hline
      				\multirow{2}{*}{All}& \textbf{BLIZZARD} vs. BL$_T$ & \textbf{2,750} (\textbf{53.51}\%) /-55 & 1,161 (22.59\%)/+32 & 1,228 (23.90\%) \\
      				\hhline{~----}
      				& BLIZZARD vs. BL & 1,947 (37.89\%)/-50 & 1,245 (24.22\%)/+30 & 1,947 (37.89)\%\\
      				\hline
      			\end{tabular}
      			\centering
      			\textbf{Preserved}=Query quality unchanged, \textbf{MRD} = Mean Rank Difference between BLIZZARD and baseline queries, \textbf{BL$_T$} = {title}, \textbf{BL} = {title + description} 
      		\end{threeparttable}
      	}
      \end{table}
  
  \begin{figure}[!t]
  	\centering
  	\includegraphics[width=3.35in ]{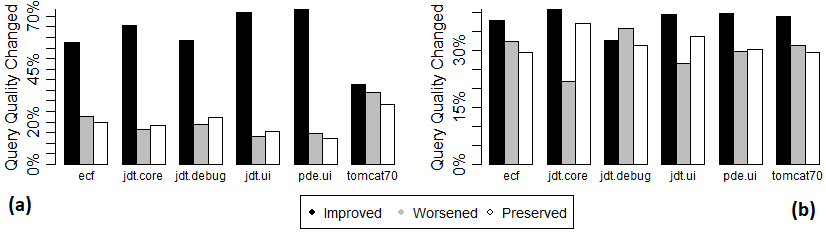}
  	\caption{Quality improvement of (a) noisy and (b) poor baseline queries by our technique--BLIZZARD}
  	\label{fig:baseline-compare-qe}
  	\vspace{-.1cm}
  \end{figure}
 
Fig. \ref{fig:compare-baseline} further demonstrates the comparative analyses between BLIZZARD and the baseline technique for various Top-K results in terms of (a) precision and (b) reciprocal rank in the bug localization.
From Fig. \ref{fig:compare-baseline}-(a), we see that precision reaches to the maximum pretty quickly (\ie\ at $K \approx 4$) for both techniques. While the baseline technique suffers from noisy (\ie\ from BR$_{ST}$) and poor (\ie\ from BR$_{NL}$) queries, BLIZZARD achieves significantly higher precision than the baseline. 
Our non-parametric statistical tests--\emph{Mann-Whitney Wilcoxon} and \emph{Cliff's Delta}--reported \emph{p-values}$<0.05$ with a \emph{large} effect size (\ie\  $0.77\le \Delta \le 1.00$).
Although the baseline precision for BR$_{PE}$ is higher, BLIZZARD offers even higher precision.
From Fig. \ref{fig:compare-baseline}-(b), we see that mean reciprocal ranks of BLIZZARD have a logarithmic shape and whereas the baseline counterparts look comparatively flat. That is, as more results from the top of the ranked list are considered, more true positives are identified by our technique than the baseline technique does. 
Statistical tests also reported strong significance (\ie\ \emph{p-values}$<$0.001) and a \emph{large} effect size (\ie\ 0.62$\le$$\Delta$$ \le$1.00) of our measures over the baseline counterparts. 
That is, BLIZZARD performs a good job in reformulating the noisy and poor queries, and such reformulations contribute to a significant improvement in the bug localization performances.    

\textbf{Answering RQ$\mathbf{_1}$(b) --Impact of Parameters and Settings:} 
We investigate the impacts of different adopted parameters -\emph{query reformulation length}, \emph{word stemming}, and \emph{retrieval engine} - upon our technique, and justify our choices. 
BLIZZARD reformulates a given query (\ie\ bug report) for bug localization, and hence, size of the reformulated query is an important parameter. Fig. \ref{fig:query-length} demonstrates how various  
reformulation lengths can affect the MAP@10 of our technique. 
We see that precision reaches the maximum for 
three report classes at different query reformulation lengths (\ie\ R$_L$). For BR$_{ST}$, we achieve the maximum precision at R$_L$=11, and for BR$_{NL}$, such maximum is detected with R$_L$ ranging between 8 and 12. On the contrary, precision increases in a logarithmic manner for BR$_{PE}$ bug reports. We investigated up to 30 reformulation terms and found the maximum precision.
Given the above empirical findings, we chose R$_L$=11 for BR$_{ST}$, R$_L$=30 for BR$_{PE}$ and R$_L$=8 for R$_{NL}$ as the adopted reformulation lengths and our choices are justified.  

We also investigate the impact of stemming and text retrieval engine on our technique. We found that stemming did not improve the performance of BLIZZARD, \ie\ reduced localization accuracy. Similar finding was reported by earlier studies as well \cite{stemming,kevic}. We also found that Lucene performs better than Indri on our dataset. Besides, Lucene has been widely used by relevant literature \cite{refoqus,trconfig,stacktrace,saner2017masud}. Given the above findings and earlier suggestions, our choices on stemming and retrieval engine are also justified. 
\vspace{-.1cm}     
\FrameSep.3em
\begin{framed}
	\noindent
	BLIZZARD outperforms baseline in accuracy, precision and reciprocal rank by 7\%--56\%, 6\%--62\% and 6\%--62\% respectively across three report groups, and our adopted parameters are also justified.    
\end{framed}
\vspace{-.1cm}

\textbf{Answering RQ$\mathbf{_2}$-Comparison with Baseline Queries:} 
While Table \ref{table:blizzard} contrasts BLIZZARD with the baseline approach for top 1 to 10 results, we further investigate how BLIZZARD performs compared to the baseline when all results of a query are considered.
We compare our queries with two baseline queries --\emph{title} (\ie\ BL$_T$), \emph{title}+\emph{description} (\ie\ BL) -- from each of the bug reports. When our query returns the first correct result at a higher position in the result list than that of corresponding baseline query, we call it \emph{query improvement} and vice versa \emph{query worsening}. When result ranks of the reformulated query and the baseline query are the same, then we call it \emph{query preserving}. 
From Table \ref{table:baseline-qe}, we see that our applied reformulations improve 59\% of the noisy queries (\ie\ BR$_{ST}$) and 39\%--56\% of the poor (\ie\ BR$_{NL}$) queries  both with $\approx$ 25\% worsening ratios. 
That is, the improvements are more than two times the worsening ratios.
Fig. \ref{fig:baseline-compare-qe} further demonstrates the potential of our reformulations where improvement, worsening and preserving ratios are plotted for each of the six subject systems. We see that noisy queries get benefited greatly from our reformulations, and on average, their query effectiveness improve up to 122 positions (\ie\ MRD of BR$_{ST}$, Table \ref{table:baseline-qe}) in the result list. Such improvement of ranks can definitely help the developers in locating the buggy files in the result list more easily. The poor queries also improve due to our reformulations significantly (\ie\ \emph{p-value}=0.004$<$0.05, \emph{Cliff's} $\Delta$=0.94 (\emph{large})), and the correct results can be found 16 positions earlier (than the baseline) in the result list starting from the top.
Quantile analysis in Table \ref{table:compare-qe} also confirms that noisy and poor queries are significantly improved by our provided reformulations. Besides, the benefits of query reformulations are also demonstrated by our findings in Table \ref{table:blizzard} and Fig. \ref{fig:compare-baseline}. 
\FrameSep.5em
\begin{framed}
	\noindent
	Our applied reformulations to the bug localization queries improve 59\% of the noisy queries and 39\%--56\% of the poor queries, and return the buggy files closer to the top of result list. Such improvements can reduce a developer's effort in locating bugs.
\end{framed}
\vspace{-.3cm}

\begin{table}[!tb]
	\centering
	\caption{Comparison with IR-Based Bug Localization Techniques}\label{table:compare}
	\resizebox{3.35in}{!}{%
		\begin{threeparttable}
			\begin{tabular}{l|l|c|c|c|c|c}
				\hline
				\textbf{RG} & \textbf{Technique}& 
				\textbf{Hit@1} & \textbf{Hit@5} & \textbf{Hit@10} & \textbf{MAP@10} & \textbf{MRR@10}\\
				\hline
				\hline
				\multirow{7}{*}{BR$_{ST}$} & BugLocator &
				 28.79\% & 55.08\% & 67.00\% & 38.49\% & 0.40 \\
				\hhline{~------}
				& BLUiR & 
				23.38\% & 44.34\% & 54.06\% & 30.96\% & 0.32\\
				\hhline{~------}
				& AmaLgam+$_{BRO}$ & 
				45.33\% & 66.97\% & 73.29\% & 52.88\% & 0.55 \\
				\hhline{~------}
				& \textbf{BLIZZARD} &  
				34.42\% & 66.28\% & \textbf{75.21}\% & 45.50\% & 0.47\\
				\hhline{~------}
				& \textbf{BLIZZARD$_{BRO}$} & \textbf{47.42}\% & \textbf{73.74}\% & \textbf{78.77}\% & \textbf{56.22}\% & \textbf{0.59}\\
				\hhline{~------}
				& AmaLgam+ & 
				50.51\% & 66.47\% & 71.66\% & 55.97\% & 0.58\\
				\hhline{~------}
				& \textbf{BLIZZARD+} & 
				\textbf{53.39}\% & \textbf{*76.12}\% & \textbf{*80.03}\% & \textbf{60.65}\% & \textbf{0.63}\\ 
				\hline
				\hline
				\multirow{7}{*}{BR$_{PE}$} & BugLocator & 
				36.25\% & 61.37\% & 70.96\% & 44.24\% & 0.47 \\
				\hhline{~------}
				& BLUiR &  
				35.54\% & 62.93\% & 72.17\% & 43.67\% & 0.47\\
				\hhline{~------}
				& AmaLgam$_{BRO}$ & 
				33.90\% & 60.48\% & 69.09\% & 42.00\% & 0.45 \\
				\hhline{~------}
				& \textbf{BLIZZARD} &
				 \textbf{*44.31}\% & \textbf{*69.48}\% & \textbf{77.84}\% & \textbf{*52.08}\% & \textbf{*0.55}\\
				\hhline{~------}
				& \textbf{BLIZZARD$_{BRO}$} & \textbf{47.16}\% & \textbf{71.26}\% & \textbf{78.25}\% & \textbf{53.69}\% & \textbf{0.57} \\
				\hhline{~------}
				& Amalgam+ & 
				 52.00\% & 68.54\% & 72.93\% & 55.80\% & 0.59 \\  
				\hhline{~------}
				& \textbf{BLIZZARD+} & 
				 \textbf{56.84}\% & \textbf{74.70}\% & \textbf{80.09}\% & \textbf{60.78}\% & \textbf{0.65}\\
				\hline
				\hline
				\multirow{7}{*}{BR$_{NL}$} & BugLocator & 
				25.11\% & 48.52\% & 59.04\% & 32.19\% & 0.35\\
				\hhline{~------}
				& BLUiR & 
				29.87\% & 56.63\% & 66.10\% & 38.07\% & 0.41 \\    
				\hhline{~------}
				& AmaLgam+$_{BRO}$ & 
				29.40\% & 56.07\% & 65.01\% & 37.74\% & 0.40 \\
				\hhline{~------}
				& \textbf{BLIZZARD} & 
				\textbf{29.16}\% & \textbf{53.78}\% & \textbf{65.21}\% & \textbf{37.62}\% & \textbf{0.40} \\
				\hhline{~------}
				& \textbf{BLIZZARD$_{BRO}$} & \textbf{35.45}\% & \textbf{58.75}\% & \textbf{69.17}\% & \textbf{42.26}\% & \textbf{0.46}\\
				\hhline{~------}
				& AmaLgam+ & 
				49.72\% & 65.42\% & 71.49\% & 52.74\% & 0.57\\
				\hhline{~------}
				& \textbf{BLIZZARD+} & 
				47.97\% & \textbf{66.24}\% & \textbf{74.49}\% & 52.12\% & 0.56 \\
				\hline
				\hline
				\multirow{7}{*}{All} & BugLocator &  
				31.85\% & 57.37\% & 67.87\% & 40.17\% & 0.43\\
				\hhline{~------}
				& BLUiR & 
				32.45\% & 59.18\% & 68.65\% & 40.82\% & 0.44\\
				\hhline{~------}
				& Amalgam+$_{BRO}$ & 
				35.03\% & 61.32\% & 69.89\% & 43.36\% & 0.46\\
				\hhline{~------}
				& \textbf{BLIZZARD} & 
				\textbf{38.58}\% & \textbf{65.08}\% & \textbf{74.52}\% & \textbf{47.13}\% & \textbf{*0.50} \\
				\hhline{~------}
				& \textbf{BLIZZARD$_{BRO}$} & \textbf{44.26}\% & \textbf{69.15}\% & \textbf{76.61}\% & \textbf{51.41}\% & \textbf{*0.55}\\
				\hhline{~------}
				& AmaLgam+ & 
				52.29\% & 68.53\% & 73.58\% & 56.03\% & 0.59 \\  
				\hhline{~------}
				& \textbf{BLIZZARD+} & 
				 \textbf{54.78}\% & \textbf{73.76}\% & \textbf{79.66}\% & \textbf{59.32}\% & \textbf{0.63}\\  
				\hline
			\end{tabular}
			\centering
			\textbf{RG}=Report Group, 
			\textbf{BRO}=Bug Report Only, \textbf{*}=Significantly higher
		\end{threeparttable}
	}
	\vspace{-.3cm}
\end{table}

\begin{table}[!t]
	\centering
	\caption{Components behind Existing IR-Based Bug Localization}\label{table:components}
	\resizebox{3.15in}{!}{%
		\begin{threeparttable}
			\begin{tabular}{l|c|c|c|c|c|c|c|c}
				\hline
				\multirow{2}{*}{\textbf{Technique}} & \multicolumn{4}{c|}{\textbf{Bug Report Only}} & 
				\multicolumn{3}{c|}{\textbf{External Resources}} &   \multirow{2}{*}{\textbf{MRR}}\\
				\hhline{~-------~}
				& \textbf{BRT} & \textbf{BRS} & \textbf{ST} & \textbf{QR} & \textbf{BRH} & \textbf{VCH} & \textbf{AH} &  \\
				\hline
				Baseline & \ding{108} & & & & & & & 0.44\\
				\hhline{---------}
				BugLocator & \ding{108} & & & & \ding{108} & & & 0.43 \\
				\hhline{---------}
				BLUiR & \ding{108} & \ding{108} & & & & & & 0.44 \\
				\hhline{---------}
				AmaLgam+$_{BRO}$ & \ding{108} & \ding{108} & \ding{108} & & & & & 0.46 \\
				\hhline{---------}  
				\textbf{BLIZZARD} & \ding{108} &  &  & \ding{108} & & & & \textbf{*0.50} \\
				\hhline{---------}
				\textbf{BLIZZARD$_{BRO}$} & \ding{108} & \ding{108} & \ding{108} & \ding{108} & & & & \textbf{*0.55}\\
				\hhline{---------}
				AmaLgam+ & \ding{108} & \ding{108} & \ding{108} &  &\ding{108} &   \ding{108} & \ding{108} & 0.59\\
				\hhline{---------}
				\textbf{BLIZZARD+} & \ding{108} & \ding{108} & \ding{108} & \ding{108} & \ding{108} & \ding{108}& \ding{108} & \textbf{0.63} \\ 
				\hline
			\end{tabular}
			\centering
			\textbf{BRT}=Bug Report Texts, \textbf{BRS}=Bug Report Structures, \textbf{ST}=Stack Traces, \textbf{QR}=Query Reformulation, \textbf{BRH}=Bug Report History, \textbf{VCH}=Version Control History, \textbf{AH}=Authoring History, \textbf{BRO}=Bug Report Only, \ding{108}=Feature used
		\end{threeparttable}
	}
\end{table}

\begin{table*}[!t]
	\centering
	\caption{Comparison of Query Effectiveness with Existing Query Reformulation Techniques}\label{table:compare-qe}
	\resizebox{6.6in}{!}{%
		\begin{threeparttable}
			\begin{tabular}{l|c|c|c|c|c|c|c|c||c|c|c|c|c|c|c||c}
				\hline
				\multirow{2}{*}{\textbf{Technique}} & \multirow{2}{*}{\textbf{RG}} &\multicolumn{7}{c||}{\textbf{Improvement}} & \multicolumn{7}{c||}{\textbf{Worsening}} & \textbf{Preserving}   \\
				\hhline{~~---------------}
				& & \#Improved & Mean & Q1 & Q2 & Q3 & Min. & Max. & \#Worsened & Mean & Q1 & Q2 & Q3 & Min. & Max. & \#Preserved\\
				\hline
				\hline
				\citet{rocchio} & \multirow{6}{*}{} & 337 (40.80\%) & 68 & 4 & 12 &  60 & 1 & 1,245 &  264 (31.96\%) & 118 & 6 & 21 & 97 & 2 & 2,824 & 225 (27.24\%) \\
				\hhline{-~---------------}
				RSV \cite{rsv} & & 218 (26.39\%) & 163 & 10 & 43 & 158 & 1 & 2,103 & 236 (28.57\%) & 198 & 17 & 71 & 245 & 2 & 2,487 & 372 (45.04\%) \\
				\hhline{-~---------------}
				\citet{sisman} & BR$_{ST}$ & 339 (41.04\%) & 66 & 4 & 12 & 53 & 1 & 1,245 & 265 (32.08\%) & 121 & 7 & 23 & 100 & 2 & 2,846 & 222 (26.88\%)\\
				\hhline{-~---------------}
				STRICT \cite{saner2017masud}& (826) & 399 (48.30\%) & 35 & 1 & 4 & 17 & 1 & 1,538 & 318 (38.50\%) & 139 & 6 & 25 & 110 & 2 & 3,066 & 109 (13.20\%)\\
				\hhline{-~---------------}
				Baseline & & & 153 & 7 & 35 & 149 & 2 & 2,221 &  & 70 & 1 & 5 & 30 & 1 & 2,469 &  \\
				\hhline{-~---------------}
				\textbf{BLIZZARD} & & \textbf{485 (58.72\%)} & \textbf{22} & \textbf{1} & \textbf{3} & \textbf{9} & \textbf{1} & \textbf{932} & \textbf{174 (21.07\%)} & \textbf{112} & \textbf{4} & \textbf{15} & \textbf{60} & \textbf{2} & 3,258 & \textbf{167 (20.22\%)}\\
				\hline 
				\hline 
				\citet{rocchio} & \multirow{6}{*}{} & 32 (2.07\%) & 33 & 4 & 8 & 19 & 1 & 365 & 24 (1.55\%) & 140 & 4 & 12 & 146 & 2 & 850 & 1,490 (96.38\%) \\
				\hhline{-~---------------}
				RSV \cite{rsv}& & 345 (22.27\%) & 112 & 3 & 9 & 38 & 1 & 6,564& 751 (48.57\%) & 105 & 7 & 23 & 81 & 2 & 2,140 & 450 (29.11\%)\\
				\hhline{-~---------------}
				\citet{sisman} & BR$_{NL}$ & 499 (32.28\%) & 59 & 2 & 6 & 26 & 1 & 2,019 & 575 (37.19\%) & 98 & 5 & 15 & 64 & 2 & 2,204 & 472 (30.47\%)\\
				\hhline{-~---------------}
				STRICT \cite{saner2017masud} & (1,546) & 467 (30.21\%) & 57 & 2 & 6 & 30 & 1 & 1,213 & 654 (42.30\%) & 112 &  5 & 18 & 63 & 2 & 4,933 & 425 (27.44\%)\\
				\hhline{-~---------------}
				Baseline & & & 91 & 5 & 15 & 57 & 2 & 2,434 &  & 61 & 2 & 8 & 30 & 1 & 1,894 & \\
				\hhline{-~---------------}\
				\textbf{BLIZZARD} & & \textbf{597 (38.62\%)} & \textbf{75} & \textbf{2} & \textbf{8} & \textbf{32} & \textbf{1} & 3,063 & \textbf{455 (29.43\%)} & \textbf{92} & \textbf{5} & \textbf{15} & \textbf{54} & \textbf{2} & 2,024 & \textbf{494 (31.95\%)}\\ 
				\hline
			\end{tabular}
		\end{threeparttable}
	}
	\vspace{-.3cm}
\end{table*}

\subsection{Comparison with Existing Techniques}\label{sec:comparison}
\textbf{Answering RQ$\mathbf{3}$ --Comparison with Existing IR-Based Bug Localization Techniques:} 
Our evaluation of BLIZZARD with four widely used performance metrics shows promising results. The comparison with the best performing baseline shows that our approach outperforms the baselines. However, in order to further gain confidence and to place our work in the literature, we also compared our approach with three IR-based bug localization techniques \cite{buglocator,bluir,versionhistoryjsep} including the state-of-the-art \cite{versionhistoryjsep}.
\citet{buglocator} first employ improved Vector Space Model (\ie\ rVSM) and bug report similarity for locating buggy source files for a new bug report. \citet{bluir} employ structured information retrieval where (1) a bug report is divided into two fields--\emph{title}, \emph{description} and a source document is divided into four fields--\emph{class}, \emph{method}, \emph{variable} and \emph{comments}, and then (2) eight similarity measures between these two groups are accumulated to rank the source document. We collect authors' implementations of both techniques for our experiments.

While the above studies use bug report contents only, the later approaches combine them \cite{bluirplus} and add more internal \cite{brtracer} or external information sources such as version control history \cite{versionhistory} and author information  \cite{versionhistoryjsep}. In the same vein, \citet{versionhistoryjsep} recently combine five internal and external information sources - similar bug report, structured IR, stack traces, version control history and bug reporter's history -- for ranking a source document, and outperform five earlier approaches which makes it the state-of-the-art in IR-based bug localization. Given that authors' implementation is not publicly available, we implement this technique ourselves by consulting with the original authors.
Since BLIZZARD does not incorporate any external information sources, to ensure a fair comparison, we also implement a variant of the state-of-the-art namely AmaLgam+$_{BRO}$. It combines bug report texts, structured IR and stack traces (\ie\ Table \ref{table:components}) for source document ranking.
 

From Table \ref{table:compare}, we see that AmaLgam+ performs the best among the existing techniques. However, its performance comes at a high cost of mining six information contents (\ie\ Table \ref{table:components}). Besides, for optimal performance, AmaLgam+ needs past bug reports, version control history and author history which might always not be available. Thus, to ensure a fair comparison, we develop two variants of our technique--BLIZZARD$_{BRO}$ and BLIZZARD+. BLIZZARD$_{BRO}$ combines query reformulation with bug report only features whereas BLIZZARD+ combines query reformulation with all ranking components of AmaLgam+ (\ie\ details in Table \ref{table:components}). We then compare both BLIZZARD and BLIZZARD$_{BRO}$ with AmaLgam+$_{BRO}$, and BLIZZARD+ with AmaLgam+ respectively. 

As shown in Table \ref{table:compare}, BLIZZARD outperforms AmaLgam+$_{BRO}$ in terms of all three metrics especially for BR$_{PE}$ reports while performing moderately high with other report groups.   
For example, BLIZZARD provides 22\% higher MRR@10 and 24\% higher MAP@10 than AmaLgam+$_{BRO}$ for BR$_{PE}$. When all report only features are complemented with appropriate query reformulations, our technique, BLIZZARD$_{BRO}$ outperforms AmaLgam+$_{BRO}$ in terms of all three metrics--Hit@K, MAP@10 and MRR@10-- with each report groups. Such findings suggest that BLIZZARD$_{BRO}$ can better exploit the available resources (\ie\ bug report contents) than the state-of-the-art variant, and returns the buggy files at relatively higher positions in the ranked list. Furthermore,  BLIZZARD+ outperforms the state-of-the-art, AmaLgam+, by introducing query reformulation paradigm. For example, BLIZZARD+ improves Hit@5 and Hit@10 over AmaLgam+ for each of the three query types, \eg\ 15\% and 12\% respectively for noisy queries (BR$_{ST}$). 
It also should be noted that none of the existing techniques is robust to all three report groups simultaneously. We overcome such issue with appropriate query reformulations, and deliver $\approx$75\%--80\% Hit@10 irrespective of the bug report quality.  
From Table \ref{table:components}, we see that BLIZZARD$_{BRO}$ provides 20\% higher MRR@10 than AmaLgam+$_{BRO}$ by consuming equal amount of resources, \ie\ bug report only. 
All these findings above suggest two important points. First, earlier studies might have failed to exploit the report contents and structures properly for bug localization. Second, query reformulation has a high potential for improving the IR-based bug localization.   
\begin{figure}[!tb]
	\centering
	\includegraphics[width=3in ]{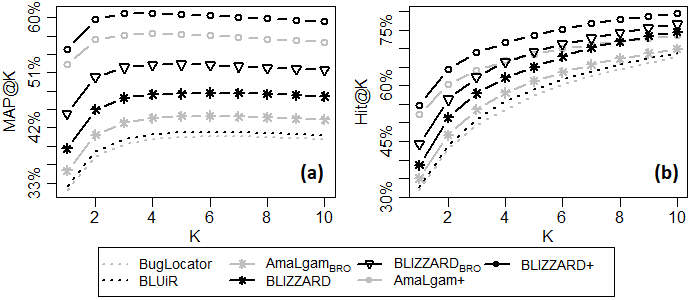}
	\caption{Comparison of (a) MAP@K and (b) Hit@K with the state-of-the-art IR-based bug localization techniques}
	\label{fig:topk-mapk-st}
	\vspace{-.5cm}
\end{figure}
\begin{figure}[!t]
	\centering
	\includegraphics[width=3.35in ]{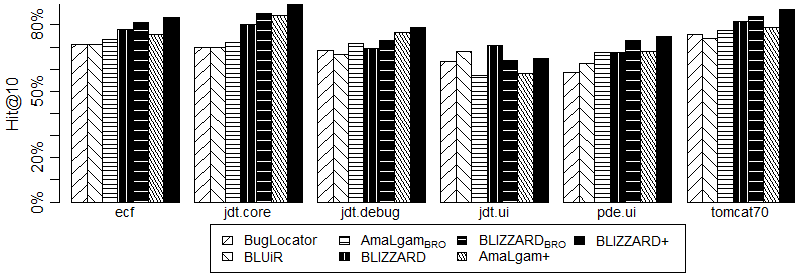}
	\caption{Comparison of Hit@10 across all subject systems}
	\label{fig:subject-acc}
\end{figure}
Fig. \ref{fig:topk-mapk-st} demonstrates a comparison of BLIZZARD with the existing techniques in terms of (a) MAP@K and (b) Hit@K for various Top-K results. Our statistical tests report that BLIZZARD, BLIZZARD$_{BRO}$ and BLIZZARD+ outperform AmaLgam+$_{BRO}$ and AmaLgam+ respectively in MAP@K by a significant margin (\ie\ \emph{p-values}$\le$0.001) and \emph{large} effect size (\ie\ 0.82$\le$$\Delta$$\le$1.00). Similar findings were also achieved for Hit@K. 

\begin{figure}[!tb]
	\centering
	\includegraphics[width=3in ]{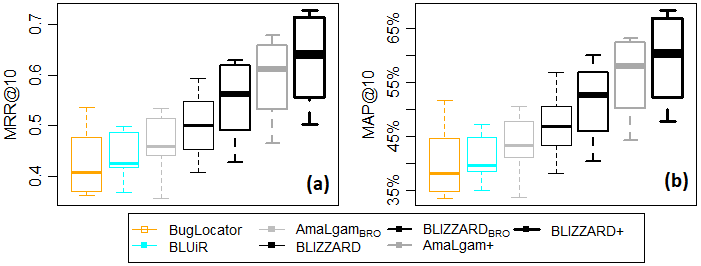}
	\caption{Comparison of (a) MRR@10 and (b) MAP@10 with existing techniques across subject systems}
	\label{fig:subject-mrrk}
\end{figure}

Fig. \ref{fig:subject-acc} and Fig. \ref{fig:subject-mrrk} focus on subject system specific performances. From Fig. \ref{fig:subject-acc}, we see that BLIZZARD outperforms AmaLgam+$_{BRO}$ with four systems in Hit@10, and falls short with two systems. However, BLIZZARD$_{BRO}$ and BLIZZARD+ outperform AmaLgam+$_{BRO}$ and AmaLgam+ respectively for all six systems. As shown in the box plots of Fig. \ref{fig:subject-mrrk}, BLIZZARD has a higher median in MRR@10 and MAP@10 than AmaLgam+$_{BRO}$ across all subject systems. AmaLgam+ improves both measures especially MAP@10. However, BLIZZARD+ provides even higher MRR@10 and MAP@10 than any of the existing techniques including the state-of-the-art. 
\vspace{-.2cm} 
\FrameSep.3em
\begin{framed}
	\noindent
	Our technique outperforms the state-of-the-art from IR-based bug localization in various dimensions. It offers 20\% higher precision and reciprocal rank than that of state-of-the-art variant (\ie\ AmaLgam+$_{BRO}$) by using only query reformulation   
	rather than costly alternatives, \eg\ mining of version control history 
\end{framed} 

\textbf{Answering RQ$\mathbf{_4}$ --Comparison with Existing Query Reformulation Techniques:} 
While we have already showed that our approach outperforms the baselines and the state-of-the-art IR-based bug localization approaches, we also wanted to further evaluate our approach in the context of query reformulation. We thus compared BLIZZARD with four query reformulation techniques \cite{rocchio,refoqus,sisman,saner2017masud} including the state-of-the-art \cite{saner2017masud} that were mostly used for concept/feature location. 
We use authors' implementation of the state-of-the-art, STRICT, and re-implement the remaining three techniques. 
We collect Query Effectiveness (\ie\ rank of the first correct result) of each of the reformulated queries provided by each technique, and compare with ours using quantile analysis. From Table \ref{table:compare-qe}, we see that 48\% of the noisy (\ie\ BR$_{ST}$) queries are improved by STRICT, and 32\% of the poor (\ie\ BR$_{NL}$) queries are improved by \citet{sisman}. Neither of these techniques considers bug report quality (\ie\ prevalence of structured information or lack thereof) and each technique applies the same reformulation strategy to all reports. On the contrary, BLIZZARD chooses appropriate reformulation based on the class of a bug report, and improves 59\% of the noisy queries and 39\% of the poor queries which are 22\% and 20\% higher respectively. When compared using quantile analysis,
we see that our quantiles are highly promising compared to the baseline. 
Our reformulations clearly improve the noisy queries, and 75\% of the improved queries return their first correct results within Top-9 (\ie\ Q$_3$=9) positions  whereas STRICT needs Top-17 positions for the same. In the case of poor queries, quantiles of BLIZZARD are comparable to that of \citeauthor{sisman}. However, BLIZZARD worsens less and preserves higher amount of the baseline queries which demonstrate its high potential.

\begin{framed}
\noindent
BLIZZARD outperforms the state-of-the-art in query reformulation using context-aware (\ie\ responsive to report quality) query reformulation.
Whatever improvements are offered to noisy and poor queries by the state-of-the-art, our technique improves 22\% more of noisy queries and 20\% more of the poor queries.  
\end{framed}
\vspace{-.1cm}

\section{Threats to Validity}\label{sec:threat}
Threats to internal validity relate to experimental errors and biases \cite{wordsim}.
Replication of existing studies and misclassification of the bug reports are possible sources of such threats. We use authors' implementation of three techniques \cite{buglocator,bluir,saner2017masud} and re-implement the remaining four. While we cannot rule out the possibility of any implementation errors, we re-implemented them by consulting with the original authors \cite{versionhistoryjsep} and their reported settings and parameters \cite{refoqus,rocchio,sisman}. 
While our technique employs appropriate regular expressions for bug report classification, they are limited in certain contexts (\eg\ ill-structured stack traces) which require limited manual analysis currently. 
More sophisticated classification approaches \cite{brcls,brclslda,defcat} could be applied in the future work. 
   
Threats to external validity relate to generalizability of a technique \cite{wordsim}. We conduct experiments using Java systems. However, since we deal with mostly structured items (\eg\ stack traces, program entities) from a bug report, our technique can be adapted to other OOP-based systems that have such structured items. 

\vspace{-.1cm}

\section{Related Work}\label{sec:related}
\textbf{Bug Localization:} 
Automated bug localization has been an active research area for over two decades \cite{bluir}. Existing studies from the literature can be roughly categorized into two broad families--\emph{spectra} based and \emph{information retrieval (IR)} based \cite{locombined,parninireval}. We deal with IR-based bug localization in this work. 
Given that spectra based techniques are costly and lack scalability \cite{parninireval,stacktrace}, several studies adopt IR-based methods such as Latent Semantic Indexing (LSI) \cite{lsiPoshyvanyk}, Latent Dirichlet Allocation (LDA) \cite{ldabug,raobug} and Vector Space Model (VSM)  \cite{buglocator,bluir,kak,stacktrace,kimtwophase,brtracer} for bug localization. They leverage the shared vocabulary between bug reports and source code entities for bug localization.
Unfortunately, as existing evidences \cite{parninireval,icse2018} suggest, they are inherently subject to the quality of bug reports. A number of recent studies complement traditional IR-based localization with spectra based analysis \cite{locombined}, machine learning \cite{hyloc,learn2rank} and mining of various repositories-- bug report history \cite{bluirplus}, version control history \cite{kak,versionhistory}, code change history \cite{blia,locus} and  bug reporter history \cite{versionhistoryjsep}. Recently, \citet{versionhistoryjsep} combine bug report contents and three external repositories, and outperform five earlier IR-based bug localization techniques \cite{buglocator,bluir,bluirplus,brtracer,versionhistory,kak} which makes it the state-of-the-art. In short, the contemporary studies advocate for combining (1) multiple localization approaches (\eg\ dynamic trace analysis \cite{locombined}, Deep learning \cite{hyloc}, learning to rank \cite{learn2rank,wembedding}) and (2) multiple external information sources with classic IR-based localization, and thus, improve the localization performances.   
However, such solutions could be costly (\ie\ multiple repository mining) and less scalable (\ie\ dependency on external information sources), and hence, could be infeasible to use in practice. 
In this work, we approach the problem differently, 
and focus on better leveraging the potential of the resources at hand (\ie\ bug report and source code) which might have been \emph{underestimated} by the earlier studies. In particular, we refine the noisy queries (\ie\ containing stack traces) 
and complement the poor queries (\ie\ lacks structured items), and offer an effective information retrieval unlike the earlier studies. Thus, issues raised by low quality bug reports \cite{parninireval} have been significantly addressed by our technique, and our experimental findings support such conjecture. We compare with three existing studies including the state-of-the-art \cite{versionhistoryjsep}, and the detailed comparison can be found in Section \ref{sec:comparison} (\ie\ RQ$_3$).

A few studies \cite{stacktrace,brtracer} analyse stack traces from a bug report for bug localization. However, they apply the trace entries to boost up source document ranking, and superfluous trace entries were not discarded from their stack traces. Learning-to-rank \cite{learn2rank,wembedding} and Deep learning \cite{hyloc} based approaches might also suffer from noisy and poor queries since they adopt classic IR without query reformulation in their document ranking. Recent studies \cite{wembedding,drewbl} employ distributional semantics of words to address limitations of VSM. Since noisy terms in the report could be an issue, our approach can complement these approaches through query reformulation.  


\textbf{Query Reformulation:} There exist several studies \cite{gayg,qperf,saner2017masud,ase2016masud,hillicse09,refoqus,kevic,wembedding,rastkar,verbose} that support concept/feature/concern location tasks using query reformulation. 
However, these approaches mostly deal with unstructured natural language texts. Thus, they might not perform well with bug reports containing excessive structured information (\eg\ stack traces), and our experimental findings also support this conjecture (Table \ref{table:compare-qe}).
\citet{sisman} first introduce query reformulation in the context of IR-based bug localization. However, their approach cannot remove noise from a query.  
Recently, \citet{observed} identify observed behaviour (OB), expected behaviour (EB) and steps to reproduce (S2R) from a bug report, and then use OB texts as a reformulated query for bug localization.
However, they only analyse unstructured texts whereas we deal with both structured and unstructured contents.
 Since we apply query reformulation, we compare with four recent query reformulation techniques employed for concept location--Rocchio \cite{rocchio}, RSV \cite{rsv}, STRICT \cite{saner2017masud} \cite{ase2017masud} and bug localization-- SCP \cite{sisman}.
The detailed comparison can be found in Section \ref{sec:comparison} (\ie\ RQ$_4$).

In short, existing IR-based techniques suffer from \emph{quality issues} of bug reports whereas traditional query reformulation techniques are \emph{not well-adapted} for the bug reports containing excessive structured information (\eg\ stack traces). 
Our work fills this \emph{gap} of the literature by incorporating context-aware (\ie\ report quality aware) query reformulation into the IR-based bug localization.
Our technique better exploits resources at hand and
delivers equal or higher performance than the state-of-the-art at a relatively lower cost.   
To the best of our knowledge, such comprehensive solution was not provided by any of the existing studies.
\vspace{-.2cm}

\section{Conclusion and Future Work}\label{sec:conclusion}
Traditional IR-based bug localization is inherently subject to the (low) quality of submitted bug reports. In this paper, we propose a novel technique that leverages the quality aspect of bug reports, incorporates context-aware query reformulation into the bug localization, and thus, overcomes such limitation. Experiments using 5,139 bug reports from six open source systems report that BLIZZARD can offer up to 62\% and 20\% higher precision   
than the best baseline technique and the state-of-the-art respectively. Our technique also improves 
22\% more of noisy queries and 20\% more of the poor queries than that of state-of-the-art.
In future, we plan to apply our learned insights and our technique to  further complex activities during debugging such as automatic bug fixing. 

\textbf{Acknowledgement:} This research was supported by Saskatchewan Innovation \& Opportunity Scholarship (2017--2018), and
the Natural Sciences and Engineering Research Council of Canada (NSERC).

\balance
 


\renewcommand*{\bibfont}{\normalsize}

\bibliographystyle{ACM-Reference-Format}
\setlength{\bibsep}{0pt plus 0.3ex}
{\normalsize
\bibliography{sigproc}  


\begin{thebibliography}{68}


\ifx \showCODEN    \undefined \def \showCODEN     #1{\unskip}     \fi
\ifx \showDOI      \undefined \def \showDOI       #1{#1}\fi
\ifx \showISBNx    \undefined \def \showISBNx     #1{\unskip}     \fi
\ifx \showISBNxiii \undefined \def \showISBNxiii  #1{\unskip}     \fi
\ifx \showISSN     \undefined \def \showISSN      #1{\unskip}     \fi
\ifx \showLCCN     \undefined \def \showLCCN      #1{\unskip}     \fi
\ifx \shownote     \undefined \def \shownote      #1{#1}          \fi
\ifx \showarticletitle \undefined \def \showarticletitle #1{#1}   \fi
\ifx \showURL      \undefined \def \showURL       {\relax}        \fi
\providecommand\bibfield[2]{#2}
\providecommand\bibinfo[2]{#2}
\providecommand\natexlab[1]{#1}
\providecommand\showeprint[2][]{arXiv:#2}

\bibitem[\protect\citeauthoryear{??}{sto}{2011}]%
        {stopword}
 \bibinfo{year}{2011}\natexlab{}.
\newblock \bibinfo{title}{Stop words}.
\newblock
\newblock
\urldef\tempurl%
\url{https://code.google.com/p/stop-words}
\showURL{%
\tempurl}
\newblock
\shownote{Accessed: June 2017.}


\bibitem[\protect\citeauthoryear{??}{key}{2015}]%
        {keyword}
 \bibinfo{year}{2015}\natexlab{}.
\newblock \bibinfo{title}{Java keywords}.
\newblock
\newblock
\urldef\tempurl%
\url{https://docs.oracle.com/javase/tutorial/java/nutsandbolts/_keywords.html}
\showURL{%
\tempurl}
\newblock
\shownote{Accessed: June 2017.}


\bibitem[\protect\citeauthoryear{??}{bli}{2018}]%
        {blizzard}
 \bibinfo{year}{2018}\natexlab{}.
\newblock \bibinfo{title}{BLIZZARD: Replication package}.
\newblock
\newblock
\urldef\tempurl%
\url{https://goo.gl/NTUqcK}
\showURL{%
\tempurl}


\bibitem[\protect\citeauthoryear{Anvik, Hiew, and Murphy}{Anvik
  et~al\mbox{.}}{2006}]%
        {whofix}
\bibfield{author}{\bibinfo{person}{J. Anvik}, \bibinfo{person}{L. Hiew}, {and}
  \bibinfo{person}{G.~C. Murphy}.} \bibinfo{year}{2006}\natexlab{}.
\newblock \showarticletitle{Who {S}hould {F}ix This {B}ug?}. In
  \bibinfo{booktitle}{\emph{Proc. ICSE}}. \bibinfo{pages}{361--370}.
\newblock


\bibitem[\protect\citeauthoryear{Ashok, Joy, Liang, Rajamani, Srinivasa, and
  Vangala}{Ashok et~al\mbox{.}}{2009}]%
        {debugadvisor}
\bibfield{author}{\bibinfo{person}{B. Ashok}, \bibinfo{person}{J. Joy},
  \bibinfo{person}{H. Liang}, \bibinfo{person}{S.~K. Rajamani},
  \bibinfo{person}{G. Srinivasa}, {and} \bibinfo{person}{V. Vangala}.}
  \bibinfo{year}{2009}\natexlab{}.
\newblock \showarticletitle{Debug{A}dvisor: {A} {R}ecommender {S}ystem for
  {D}ebugging}. In \bibinfo{booktitle}{\emph{Proc. ESEC/FSE}}.
  \bibinfo{pages}{373--382}.
\newblock


\bibitem[\protect\citeauthoryear{Bachmann and Bernstein}{Bachmann and
  Bernstein}{2009}]%
        {bugid}
\bibfield{author}{\bibinfo{person}{A. Bachmann} {and} \bibinfo{person}{A.
  Bernstein}.} \bibinfo{year}{2009}\natexlab{}.
\newblock \showarticletitle{Software Process Data Quality and Characteristics:
  A Historical View on Open and Closed Source Projects}. In
  \bibinfo{booktitle}{\emph{Proc. IWPSE}}. \bibinfo{pages}{119--128}.
\newblock


\bibitem[\protect\citeauthoryear{Bassett and Kraft}{Bassett and Kraft}{2013}]%
        {twkraft}
\bibfield{author}{\bibinfo{person}{B. Bassett} {and} \bibinfo{person}{N.~A.
  Kraft}.} \bibinfo{year}{2013}\natexlab{}.
\newblock \showarticletitle{Structural information based term weighting in text
  retrieval for feature location}. In \bibinfo{booktitle}{\emph{Proc. ICPC}}.
  \bibinfo{pages}{133--141}.
\newblock


\bibitem[\protect\citeauthoryear{Bettenburg, Just, Schr\"{o}ter, Weiss,
  Premraj, and Zimmermann}{Bettenburg et~al\mbox{.}}{2008a}]%
        {goodbugreport}
\bibfield{author}{\bibinfo{person}{N. Bettenburg}, \bibinfo{person}{S. Just},
  \bibinfo{person}{A. Schr\"{o}ter}, \bibinfo{person}{C. Weiss},
  \bibinfo{person}{R. Premraj}, {and} \bibinfo{person}{T. Zimmermann}.}
  \bibinfo{year}{2008}\natexlab{a}.
\newblock \showarticletitle{What Makes a Good Bug Report?}. In
  \bibinfo{booktitle}{\emph{Proc. FSE}}. \bibinfo{pages}{308--318}.
\newblock


\bibitem[\protect\citeauthoryear{Bettenburg, Premraj, Zimmermann, and
  Kim}{Bettenburg et~al\mbox{.}}{2008b}]%
        {structcls}
\bibfield{author}{\bibinfo{person}{N. Bettenburg}, \bibinfo{person}{R.
  Premraj}, \bibinfo{person}{T. Zimmermann}, {and} \bibinfo{person}{S. Kim}.}
  \bibinfo{year}{2008}\natexlab{b}.
\newblock \showarticletitle{Extracting Structural Information from Bug
  Reports}. In \bibinfo{booktitle}{\emph{Proc. MSR}}. \bibinfo{pages}{27--30}.
\newblock


\bibitem[\protect\citeauthoryear{Blanco and Lioma}{Blanco and Lioma}{2012}]%
        {blanco}
\bibfield{author}{\bibinfo{person}{R. Blanco} {and} \bibinfo{person}{C.
  Lioma}.} \bibinfo{year}{2012}\natexlab{}.
\newblock \showarticletitle{Graph-based {T}erm {W}eighting for {I}nformation
  {R}etrieval}.
\newblock \bibinfo{journal}{\emph{Inf. Retr.}} \bibinfo{volume}{15},
  \bibinfo{number}{1} (\bibinfo{year}{2012}), \bibinfo{pages}{54--92}.
\newblock


\bibitem[\protect\citeauthoryear{Brin and Page}{Brin and Page}{1998}]%
        {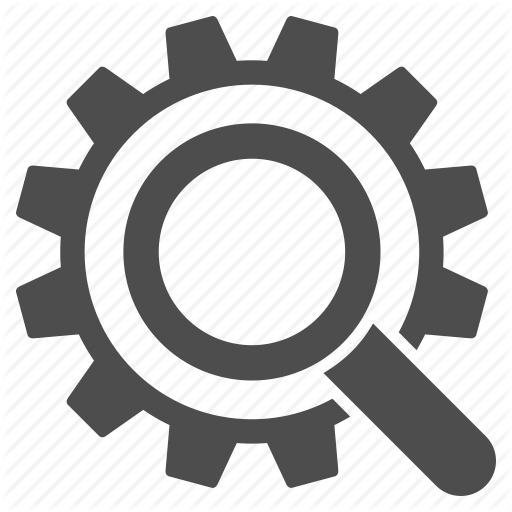}
\bibfield{author}{\bibinfo{person}{S. Brin} {and} \bibinfo{person}{L. Page}.}
  \bibinfo{year}{1998}\natexlab{}.
\newblock \showarticletitle{The {A}natomy of a {L}arge-{S}cale {H}ypertextual
  {W}eb {S}earch {E}ngine}.
\newblock \bibinfo{journal}{\emph{Comput. Netw. ISDN Syst.}}
  \bibinfo{volume}{30}, \bibinfo{number}{1-7} (\bibinfo{year}{1998}),
  \bibinfo{pages}{107--117}.
\newblock


\bibitem[\protect\citeauthoryear{Carpineto and Romano}{Carpineto and
  Romano}{2012}]%
        {qsurvey}
\bibfield{author}{\bibinfo{person}{C. Carpineto} {and} \bibinfo{person}{G.
  Romano}.} \bibinfo{year}{2012}\natexlab{}.
\newblock \showarticletitle{A Survey of Automatic Query Expansion in
  Information Retrieval}.
\newblock \bibinfo{journal}{\emph{ACM Comput. Surv.}} \bibinfo{volume}{44},
  \bibinfo{number}{1} (\bibinfo{year}{2012}), \bibinfo{pages}{1:1--1:50}.
\newblock


\bibitem[\protect\citeauthoryear{Chaparro, Florez, and Marcus}{Chaparro
  et~al\mbox{.}}{2017}]%
        {observed}
\bibfield{author}{\bibinfo{person}{O. Chaparro}, \bibinfo{person}{J.~M.
  Florez}, {and} \bibinfo{person}{A Marcus}.} \bibinfo{year}{2017}\natexlab{}.
\newblock \showarticletitle{Using Observed Behavior to Reformulate Queries
  during Text Retrieval-based Bug Localization}. In
  \bibinfo{booktitle}{\emph{Proc. ICSME}}. \bibinfo{pages}{to appear}.
\newblock


\bibitem[\protect\citeauthoryear{Chaparro and Marcus}{Chaparro and
  Marcus}{2016}]%
        {verbose}
\bibfield{author}{\bibinfo{person}{O. Chaparro} {and} \bibinfo{person}{A.
  Marcus}.} \bibinfo{year}{2016}\natexlab{}.
\newblock \showarticletitle{On the Reduction of Verbose Queries in Text
  Retrieval Based Software Maintenance}. In \bibinfo{booktitle}{\emph{Proc.
  ICSE-C}}. \bibinfo{pages}{716--718}.
\newblock


\bibitem[\protect\citeauthoryear{Chen and Kim}{Chen and Kim}{2015}]%
        {crowddebug}
\bibfield{author}{\bibinfo{person}{F. Chen} {and} \bibinfo{person}{S. Kim}.}
  \bibinfo{year}{2015}\natexlab{}.
\newblock \showarticletitle{Crowd {D}ebugging}. In
  \bibinfo{booktitle}{\emph{Proc. ESEC/FSE}}. \bibinfo{pages}{320--332}.
\newblock


\bibitem[\protect\citeauthoryear{Cordeiro, Antunes, and Gomes}{Cordeiro
  et~al\mbox{.}}{2012}]%
        {context}
\bibfield{author}{\bibinfo{person}{J. Cordeiro}, \bibinfo{person}{B. Antunes},
  {and} \bibinfo{person}{P. Gomes}.} \bibinfo{year}{2012}\natexlab{}.
\newblock \showarticletitle{{C}ontext-based {R}ecommendation to {S}upport
  {P}roblem {S}olving in {S}oftware {D}evelopment}. In
  \bibinfo{booktitle}{\emph{Proc. RSSE}}. \bibinfo{pages}{85 --89}.
\newblock


\bibitem[\protect\citeauthoryear{Enslen, Hill, Pollock, and
  Vijay-Shanker}{Enslen et~al\mbox{.}}{2009}]%
        {samurai}
\bibfield{author}{\bibinfo{person}{E. Enslen}, \bibinfo{person}{E. Hill},
  \bibinfo{person}{L. Pollock}, {and} \bibinfo{person}{K. Vijay-Shanker}.}
  \bibinfo{year}{2009}\natexlab{}.
\newblock \showarticletitle{Mining source code to automatically split
  identifiers for software analysis}. In \bibinfo{booktitle}{\emph{Proc. MSR}}.
  \bibinfo{pages}{71--80}.
\newblock


\bibitem[\protect\citeauthoryear{Gay, Haiduc, Marcus, and Menzies}{Gay
  et~al\mbox{.}}{2009}]%
        {gayg}
\bibfield{author}{\bibinfo{person}{G. Gay}, \bibinfo{person}{S. Haiduc},
  \bibinfo{person}{A. Marcus}, {and} \bibinfo{person}{T. Menzies}.}
  \bibinfo{year}{2009}\natexlab{}.
\newblock \showarticletitle{On the {U}se of {R}elevance {F}eedback in
  {IR}-based {C}oncept {L}ocation}. In \bibinfo{booktitle}{\emph{Proc. ICSM}}.
  \bibinfo{pages}{351--360}.
\newblock


\bibitem[\protect\citeauthoryear{Gu, Barr, Schleck, and Su}{Gu
  et~al\mbox{.}}{2012}]%
        {reuse}
\bibfield{author}{\bibinfo{person}{Z. Gu}, \bibinfo{person}{E.T. Barr},
  \bibinfo{person}{D. Schleck}, {and} \bibinfo{person}{Z. Su}.}
  \bibinfo{year}{2012}\natexlab{}.
\newblock \showarticletitle{Reusing {D}ebugging {K}nowledge via {T}race-based
  {B}ug {S}earch}. In \bibinfo{booktitle}{\emph{Proc. OOPSLA}}.
  \bibinfo{pages}{927--942}.
\newblock


\bibitem[\protect\citeauthoryear{Haiduc, Bavota, Marcus, Oliveto, De~Lucia, and
  Menzies}{Haiduc et~al\mbox{.}}{2013}]%
        {refoqus}
\bibfield{author}{\bibinfo{person}{S. Haiduc}, \bibinfo{person}{G. Bavota},
  \bibinfo{person}{A. Marcus}, \bibinfo{person}{R. Oliveto},
  \bibinfo{person}{A. De~Lucia}, {and} \bibinfo{person}{T. Menzies}.}
  \bibinfo{year}{2013}\natexlab{}.
\newblock \showarticletitle{Automatic {Q}uery {R}eformulations for {T}ext
  {R}etrieval in {S}oftware {E}ngineering}. In \bibinfo{booktitle}{\emph{Proc.
  ICSE}}. \bibinfo{pages}{842--851}.
\newblock


\bibitem[\protect\citeauthoryear{Haiduc, Bavota, Oliveto, De~Lucia, and
  Marcus}{Haiduc et~al\mbox{.}}{2012}]%
        {qperf}
\bibfield{author}{\bibinfo{person}{S. Haiduc}, \bibinfo{person}{G. Bavota},
  \bibinfo{person}{R. Oliveto}, \bibinfo{person}{A. De~Lucia}, {and}
  \bibinfo{person}{A. Marcus}.} \bibinfo{year}{2012}\natexlab{}.
\newblock \showarticletitle{Automatic Query Performance Assessment During the
  Retrieval of Software Artifacts}. In \bibinfo{booktitle}{\emph{Proc. ASE}}.
  \bibinfo{pages}{90--99}.
\newblock


\bibitem[\protect\citeauthoryear{Hill, Pollock, and Vijay-Shanker}{Hill
  et~al\mbox{.}}{2009}]%
        {hillicse09}
\bibfield{author}{\bibinfo{person}{E. Hill}, \bibinfo{person}{L. Pollock},
  {and} \bibinfo{person}{K. Vijay-Shanker}.} \bibinfo{year}{2009}\natexlab{}.
\newblock \showarticletitle{Automatically Capturing Source Code Context of
  NL-queries for Software Maintenance and Reuse}. In
  \bibinfo{booktitle}{\emph{Proc. ICSE}}. \bibinfo{pages}{232--242}.
\newblock


\bibitem[\protect\citeauthoryear{Hill, Rao, and Kak}{Hill
  et~al\mbox{.}}{2012}]%
        {stemming}
\bibfield{author}{\bibinfo{person}{E Hill}, \bibinfo{person}{S Rao}, {and}
  \bibinfo{person}{A Kak}.} \bibinfo{year}{2012}\natexlab{}.
\newblock \showarticletitle{{On the {U}se of {S}temming for {C}oncern
  {L}ocation and {B}ug {L}ocalization in {J}ava}}. In
  \bibinfo{booktitle}{\emph{Proc. SCAM}}. \bibinfo{pages}{184--193}.
\newblock


\bibitem[\protect\citeauthoryear{Jespersen}{Jespersen}{1929}]%
        {jespersen}
\bibfield{author}{\bibinfo{person}{O. Jespersen}.}
  \bibinfo{year}{1929}\natexlab{}.
\newblock \showarticletitle{{The Philosophy of Grammar}}.
\newblock  (\bibinfo{year}{1929}).
\newblock


\bibitem[\protect\citeauthoryear{Jones}{Jones}{1972}]%
        {tfidf}
\bibfield{author}{\bibinfo{person}{K~S Jones}.}
  \bibinfo{year}{1972}\natexlab{}.
\newblock \showarticletitle{{A {S}tatistical {I}nterpretation Of {T}erm
  {S}pecificity And {I}ts {A}pplication In {R}etrieval}}.
\newblock \bibinfo{journal}{\emph{J. Doc.}} \bibinfo{volume}{28},
  \bibinfo{number}{1} (\bibinfo{year}{1972}), \bibinfo{pages}{11--21}.
\newblock


\bibitem[\protect\citeauthoryear{Kevic and Fritz}{Kevic and Fritz}{2014}]%
        {kevic}
\bibfield{author}{\bibinfo{person}{K. Kevic} {and} \bibinfo{person}{T. Fritz}.}
  \bibinfo{year}{2014}\natexlab{}.
\newblock \showarticletitle{Automatic {S}earch {T}erm {I}dentification for
  {C}hange {T}asks}. In \bibinfo{booktitle}{\emph{Proc. ICSE}}.
  \bibinfo{pages}{468--471}.
\newblock


\bibitem[\protect\citeauthoryear{Kim, Tao, Kim, and Zeller}{Kim
  et~al\mbox{.}}{2013}]%
        {kimtwophase}
\bibfield{author}{\bibinfo{person}{D. Kim}, \bibinfo{person}{Y. Tao},
  \bibinfo{person}{S. Kim}, {and} \bibinfo{person}{A. Zeller}.}
  \bibinfo{year}{2013}\natexlab{}.
\newblock \showarticletitle{Where Should We Fix This Bug? A Two-Phase
  Recommendation Model}.
\newblock \bibinfo{journal}{\emph{TSE}} \bibinfo{volume}{39},
  \bibinfo{number}{11} (\bibinfo{year}{2013}), \bibinfo{pages}{1597--1610}.
\newblock


\bibitem[\protect\citeauthoryear{Lam, Nguyen, Nguyen, and Nguyen}{Lam
  et~al\mbox{.}}{2017}]%
        {hyloc}
\bibfield{author}{\bibinfo{person}{A.~N. Lam}, \bibinfo{person}{A.~T. Nguyen},
  \bibinfo{person}{H.~A. Nguyen}, {and} \bibinfo{person}{T.~N. Nguyen}.}
  \bibinfo{year}{2017}\natexlab{}.
\newblock \showarticletitle{Bug Localization with Combination of Deep Learning
  and Information Retrieval}. In \bibinfo{booktitle}{\emph{Proc. ICPC}}.
  \bibinfo{pages}{218--229}.
\newblock


\bibitem[\protect\citeauthoryear{Le, Oentaryo, and Lo}{Le
  et~al\mbox{.}}{2015}]%
        {locombined}
\bibfield{author}{\bibinfo{person}{Tien-Duy~B. Le}, \bibinfo{person}{R.~J.
  Oentaryo}, {and} \bibinfo{person}{D. Lo}.} \bibinfo{year}{2015}\natexlab{}.
\newblock \showarticletitle{Information Retrieval and Spectrum Based Bug
  Localization: Better Together}. In \bibinfo{booktitle}{\emph{Proc.
  ESEC/FSE}}. \bibinfo{pages}{579--590}.
\newblock


\bibitem[\protect\citeauthoryear{Mihalcea}{Mihalcea}{2005}]%
        {wordsense}
\bibfield{author}{\bibinfo{person}{R. Mihalcea}.}
  \bibinfo{year}{2005}\natexlab{}.
\newblock \showarticletitle{Unsupervised Large-vocabulary Word Sense
  Disambiguation with Graph-based Algorithms for Sequence Data Labeling}. In
  \bibinfo{booktitle}{\emph{Proc. HLT}}. \bibinfo{pages}{411--418}.
\newblock


\bibitem[\protect\citeauthoryear{Mihalcea and Tarau}{Mihalcea and
  Tarau}{2004}]%
        {rada}
\bibfield{author}{\bibinfo{person}{R. Mihalcea} {and} \bibinfo{person}{P.
  Tarau}.} \bibinfo{year}{2004}\natexlab{}.
\newblock \showarticletitle{TextRank: {B}ringing {O}rder into {T}exts}. In
  \bibinfo{booktitle}{\emph{Proc. EMNLP}}. \bibinfo{pages}{404--411}.
\newblock


\bibitem[\protect\citeauthoryear{Mikolov, Sutskever, Chen, Corrado, and
  Dean}{Mikolov et~al\mbox{.}}{2013}]%
        {w2vec}
\bibfield{author}{\bibinfo{person}{T. Mikolov}, \bibinfo{person}{I. Sutskever},
  \bibinfo{person}{K. Chen}, \bibinfo{person}{G.~S. Corrado}, {and}
  \bibinfo{person}{J. Dean}.} \bibinfo{year}{2013}\natexlab{}.
\newblock \showarticletitle{Distributed Representations of Words and Phrases
  and their Compositionality}. In \bibinfo{booktitle}{\emph{Proc. NIPS}}.
  \bibinfo{pages}{3111--3119}.
\newblock


\bibitem[\protect\citeauthoryear{Moreno, Bavota, Haiduc, Di~Penta, Oliveto,
  Russo, and Marcus}{Moreno et~al\mbox{.}}{2015}]%
        {trconfig}
\bibfield{author}{\bibinfo{person}{L. Moreno}, \bibinfo{person}{G. Bavota},
  \bibinfo{person}{S. Haiduc}, \bibinfo{person}{M. Di~Penta},
  \bibinfo{person}{R. Oliveto}, \bibinfo{person}{B. Russo}, {and}
  \bibinfo{person}{A. Marcus}.} \bibinfo{year}{2015}\natexlab{}.
\newblock \showarticletitle{Query-based Configuration of Text Retrieval
  Solutions for Software Engineering Tasks}. In \bibinfo{booktitle}{\emph{Proc.
  ESEC/FSE}}. \bibinfo{pages}{567--578}.
\newblock


\bibitem[\protect\citeauthoryear{Moreno, Treadway, Marcus, and Shen}{Moreno
  et~al\mbox{.}}{2014}]%
        {stacktrace}
\bibfield{author}{\bibinfo{person}{L. Moreno}, \bibinfo{person}{J.~J.
  Treadway}, \bibinfo{person}{A. Marcus}, {and} \bibinfo{person}{W. Shen}.}
  \bibinfo{year}{2014}\natexlab{}.
\newblock \showarticletitle{On the Use of Stack Traces to Improve Text
  Retrieval-Based Bug Localization}. In \bibinfo{booktitle}{\emph{Proc.
  ICSME}}. \bibinfo{pages}{151--160}.
\newblock


\bibitem[\protect\citeauthoryear{Mujumdar, Kallenbach, Liu, and
  Hartmann}{Mujumdar et~al\mbox{.}}{2011}]%
        {majumdar}
\bibfield{author}{\bibinfo{person}{D. Mujumdar}, \bibinfo{person}{M.
  Kallenbach}, \bibinfo{person}{B. Liu}, {and} \bibinfo{person}{B. Hartmann}.}
  \bibinfo{year}{2011}\natexlab{}.
\newblock \showarticletitle{Crowdsourcing {S}uggestions to {P}rogramming
  {P}roblems for {D}ynamic {W}eb {D}evelopment {L}anguages}. In
  \bibinfo{booktitle}{\emph{Proc. CHI}}. \bibinfo{pages}{1525--1530}.
\newblock


\bibitem[\protect\citeauthoryear{Nguyen, Nguyen, Al-Kofahi, Nguyen, and
  Nguyen}{Nguyen et~al\mbox{.}}{2011}]%
        {ldabug}
\bibfield{author}{\bibinfo{person}{A.~T. Nguyen}, \bibinfo{person}{T.~T.
  Nguyen}, \bibinfo{person}{J. Al-Kofahi}, \bibinfo{person}{H.~V. Nguyen},
  {and} \bibinfo{person}{T.~N. Nguyen}.} \bibinfo{year}{2011}\natexlab{}.
\newblock \showarticletitle{A Topic-based Approach for Narrowing the Search
  Space of Buggy Files from a Bug Report}. In \bibinfo{booktitle}{\emph{Proc.
  ASE}}. \bibinfo{pages}{263--272}.
\newblock


\bibitem[\protect\citeauthoryear{Parnin and Orso}{Parnin and Orso}{2011}]%
        {parnin}
\bibfield{author}{\bibinfo{person}{C. Parnin} {and} \bibinfo{person}{A. Orso}.}
  \bibinfo{year}{2011}\natexlab{}.
\newblock \showarticletitle{{Are Automated Debugging Techniques Actually
  Helping Programmers?}}. In \bibinfo{booktitle}{\emph{Proc. ISSTA}}.
  \bibinfo{pages}{199--209}.
\newblock


\bibitem[\protect\citeauthoryear{Pingclasai, Hata, and i.~Matsumoto}{Pingclasai
  et~al\mbox{.}}{2013}]%
        {brclslda}
\bibfield{author}{\bibinfo{person}{N. Pingclasai}, \bibinfo{person}{H. Hata},
  {and} \bibinfo{person}{K. i. Matsumoto}.} \bibinfo{year}{2013}\natexlab{}.
\newblock \showarticletitle{Classifying Bug Reports to Bugs and Other Requests
  Using Topic Modeling}. In \bibinfo{booktitle}{\emph{Proc. APSEC}},
  Vol.~\bibinfo{volume}{2}. \bibinfo{pages}{13--18}.
\newblock


\bibitem[\protect\citeauthoryear{Poshyvanyk, Gueheneuc, Marcus, Antoniol, and
  Rajlich}{Poshyvanyk et~al\mbox{.}}{2007}]%
        {lsiPoshyvanyk}
\bibfield{author}{\bibinfo{person}{D. Poshyvanyk}, \bibinfo{person}{Y.~G.
  Gueheneuc}, \bibinfo{person}{A. Marcus}, \bibinfo{person}{G. Antoniol}, {and}
  \bibinfo{person}{V. Rajlich}.} \bibinfo{year}{2007}\natexlab{}.
\newblock \showarticletitle{Feature Location Using Probabilistic Ranking of
  Methods Based on Execution Scenarios and Information Retrieval}.
\newblock \bibinfo{journal}{\emph{TSE}} \bibinfo{volume}{33},
  \bibinfo{number}{6} (\bibinfo{year}{2007}), \bibinfo{pages}{420--432}.
\newblock


\bibitem[\protect\citeauthoryear{Rahman and Roy}{Rahman and Roy}{2016}]%
        {ase2016masud}
\bibfield{author}{\bibinfo{person}{M.~M. Rahman} {and} \bibinfo{person}{C.~K.
  Roy}.} \bibinfo{year}{2016}\natexlab{}.
\newblock \showarticletitle{{QUICKAR: Automatic Query Reformulation for Concept
  Location Using Crowdsourced Knowledge}}. In \bibinfo{booktitle}{\emph{Proc.
  ASE}}. \bibinfo{pages}{220--225}.
\newblock


\bibitem[\protect\citeauthoryear{Rahman and Roy}{Rahman and Roy}{2017a}]%
        {ase2017masud}
\bibfield{author}{\bibinfo{person}{M.~M. Rahman} {and} \bibinfo{person}{C.~K.
  Roy}.} \bibinfo{year}{2017}\natexlab{a}.
\newblock \showarticletitle{Improved Query Reformulation for Concept Location
  using CodeRank and Document Structures}. In \bibinfo{booktitle}{\emph{Proc.
  ASE}}. \bibinfo{pages}{428--439}.
\newblock


\bibitem[\protect\citeauthoryear{Rahman and Roy}{Rahman and Roy}{2017b}]%
        {saner2017masud}
\bibfield{author}{\bibinfo{person}{M.~M. Rahman} {and} \bibinfo{person}{C.~K.
  Roy}.} \bibinfo{year}{2017}\natexlab{b}.
\newblock \showarticletitle{{STRICT}: {Information Retrieval Based Search Term
  Identification for Concept Location}}. In \bibinfo{booktitle}{\emph{Proc.
  SANER}}. \bibinfo{pages}{79--90}.
\newblock


\bibitem[\protect\citeauthoryear{Rahman and Roy}{Rahman and Roy}{2018}]%
        {icse2018}
\bibfield{author}{\bibinfo{person}{M.~M. Rahman} {and} \bibinfo{person}{C.~K.
  Roy}.} \bibinfo{year}{2018}\natexlab{}.
\newblock \showarticletitle{Improving Bug Localization with Report Quality
  Dynamics and Query Reformulation}. In \bibinfo{booktitle}{\emph{Proc.
  ICSE-C}}. \bibinfo{pages}{348--349}.
\newblock


\bibitem[\protect\citeauthoryear{Rao and Kak}{Rao and Kak}{2011}]%
        {raobug}
\bibfield{author}{\bibinfo{person}{S. Rao} {and} \bibinfo{person}{A. Kak}.}
  \bibinfo{year}{2011}\natexlab{}.
\newblock \showarticletitle{Retrieval from Software Libraries for Bug
  Localization: A Comparative Study of Generic and Composite Text Models}. In
  \bibinfo{booktitle}{\emph{Proc. MSR}}. \bibinfo{pages}{43--52}.
\newblock


\bibitem[\protect\citeauthoryear{Rastkar, Murphy, and Murray}{Rastkar
  et~al\mbox{.}}{2010}]%
        {rastkar}
\bibfield{author}{\bibinfo{person}{S. Rastkar}, \bibinfo{person}{G.~C. Murphy},
  {and} \bibinfo{person}{G. Murray}.} \bibinfo{year}{2010}\natexlab{}.
\newblock \showarticletitle{Summarizing Software Artifacts: A Case Study of Bug
  Reports}. In \bibinfo{booktitle}{\emph{Proc. ICSE}}.
  \bibinfo{pages}{505--514}.
\newblock


\bibitem[\protect\citeauthoryear{Rigby and Robillard}{Rigby and
  Robillard}{2013}]%
        {rigby}
\bibfield{author}{\bibinfo{person}{P.~C. Rigby} {and} \bibinfo{person}{M.P.
  Robillard}.} \bibinfo{year}{2013}\natexlab{}.
\newblock \showarticletitle{{Discovering {E}ssential {C}ode {E}lements in
  {I}nformal {D}ocumentation}}. In \bibinfo{booktitle}{\emph{Proc. ICSE}}.
  \bibinfo{pages}{832--841}.
\newblock


\bibitem[\protect\citeauthoryear{Robertson}{Robertson}{1991}]%
        {rsv}
\bibfield{author}{\bibinfo{person}{S.~E. Robertson}.}
  \bibinfo{year}{1991}\natexlab{}.
\newblock \showarticletitle{On Term Selection for Query Expansion}.
\newblock \bibinfo{journal}{\emph{J. Doc.}} \bibinfo{volume}{46},
  \bibinfo{number}{4} (\bibinfo{year}{1991}), \bibinfo{pages}{359--364}.
\newblock


\bibitem[\protect\citeauthoryear{Rocchio}{Rocchio}{[n. d.]}]%
        {rocchio}
\bibfield{author}{\bibinfo{person}{J.J. Rocchio}.} \bibinfo{year}{[n.
  d.]}\natexlab{}.
\newblock \bibinfo{booktitle}{\emph{The SMART Retrieval System---Experiments in
  Automatic Document Processing}}.
\newblock \bibinfo{publisher}{Prentice-Hall, Inc.} 313--323 pages.
\newblock


\bibitem[\protect\citeauthoryear{Saha, Lawall, Khurshid, and Perry}{Saha
  et~al\mbox{.}}{2014}]%
        {bluirplus}
\bibfield{author}{\bibinfo{person}{R.~K. Saha}, \bibinfo{person}{J. Lawall},
  \bibinfo{person}{S. Khurshid}, {and} \bibinfo{person}{D.~E. Perry}.}
  \bibinfo{year}{2014}\natexlab{}.
\newblock \showarticletitle{On the Effectiveness of Information Retrieval Based
  Bug Localization for C Programs}. In \bibinfo{booktitle}{\emph{Proc. ICSME}}.
  \bibinfo{pages}{161--170}.
\newblock


\bibitem[\protect\citeauthoryear{Saha, Lease, Khurshid, and Perry}{Saha
  et~al\mbox{.}}{2013}]%
        {bluir}
\bibfield{author}{\bibinfo{person}{R.~K. Saha}, \bibinfo{person}{M. Lease},
  \bibinfo{person}{S. Khurshid}, {and} \bibinfo{person}{D.~E. Perry}.}
  \bibinfo{year}{2013}\natexlab{}.
\newblock \showarticletitle{Improving bug localization using structured
  information retrieval}. In \bibinfo{booktitle}{\emph{Proc. ASE}}.
  \bibinfo{pages}{345--355}.
\newblock


\bibitem[\protect\citeauthoryear{Sisman and Kak}{Sisman and Kak}{2012}]%
        {kak}
\bibfield{author}{\bibinfo{person}{B. Sisman} {and} \bibinfo{person}{A.~C.
  Kak}.} \bibinfo{year}{2012}\natexlab{}.
\newblock \showarticletitle{Incorporating Version Histories in Information
  Retrieval Based Bug Localization}. In \bibinfo{booktitle}{\emph{Proc. MSR}}.
  \bibinfo{pages}{50--59}.
\newblock


\bibitem[\protect\citeauthoryear{Sisman and Kak}{Sisman and Kak}{2013}]%
        {sisman}
\bibfield{author}{\bibinfo{person}{B. Sisman} {and} \bibinfo{person}{A.~C.
  Kak}.} \bibinfo{year}{2013}\natexlab{}.
\newblock \showarticletitle{Assisting code search with automatic Query
  Reformulation for bug localization}. In \bibinfo{booktitle}{\emph{Proc.
  MSR}}. \bibinfo{pages}{309--318}.
\newblock


\bibitem[\protect\citeauthoryear{Thung, Lo, and Jiang}{Thung
  et~al\mbox{.}}{2012}]%
        {defcat}
\bibfield{author}{\bibinfo{person}{F. Thung}, \bibinfo{person}{D. Lo}, {and}
  \bibinfo{person}{L. Jiang}.} \bibinfo{year}{2012}\natexlab{}.
\newblock \showarticletitle{Automatic Defect Categorization}. In
  \bibinfo{booktitle}{\emph{Proc. WCRE}}. \bibinfo{pages}{205--214}.
\newblock


\bibitem[\protect\citeauthoryear{Toutanova, Klein, Manning, and
  Singer}{Toutanova et~al\mbox{.}}{2003}]%
        {postagger}
\bibfield{author}{\bibinfo{person}{K. Toutanova}, \bibinfo{person}{D. Klein},
  \bibinfo{person}{C.D. Manning}, {and} \bibinfo{person}{Y. Singer}.}
  \bibinfo{year}{2003}\natexlab{}.
\newblock \showarticletitle{{Feature-Rich Part-of-Speech Tagging with a Cyclic
  Dependency Network}}. In \bibinfo{booktitle}{\emph{Proc. HLT-NAACL}}.
  \bibinfo{pages}{252--259}.
\newblock


\bibitem[\protect\citeauthoryear{Uneno, Mizuno, and Choi}{Uneno
  et~al\mbox{.}}{2016}]%
        {drewbl}
\bibfield{author}{\bibinfo{person}{Y. Uneno}, \bibinfo{person}{O. Mizuno},
  {and} \bibinfo{person}{E.~H. Choi}.} \bibinfo{year}{2016}\natexlab{}.
\newblock \showarticletitle{Using a Distributed Representation of Words in
  Localizing Relevant Files for Bug Reports}. In
  \bibinfo{booktitle}{\emph{Proc. QRS}}. \bibinfo{pages}{183--190}.
\newblock


\bibitem[\protect\citeauthoryear{Wang, Parnin, and Orso}{Wang
  et~al\mbox{.}}{2015}]%
        {parninireval}
\bibfield{author}{\bibinfo{person}{Q. Wang}, \bibinfo{person}{C. Parnin}, {and}
  \bibinfo{person}{A. Orso}.} \bibinfo{year}{2015}\natexlab{}.
\newblock \showarticletitle{Evaluating the Usefulness of IR-based Fault
  Localization Techniques}. In \bibinfo{booktitle}{\emph{Proc. ISSTA}}.
  \bibinfo{pages}{1--11}.
\newblock


\bibitem[\protect\citeauthoryear{Wang and Lo}{Wang and Lo}{2014}]%
        {versionhistory}
\bibfield{author}{\bibinfo{person}{S. Wang} {and} \bibinfo{person}{D. Lo}.}
  \bibinfo{year}{2014}\natexlab{}.
\newblock \showarticletitle{Version History, Similar Report, and Structure:
  Putting Them Together for Improved Bug Localization}. In
  \bibinfo{booktitle}{\emph{Proc. ICPC}}. \bibinfo{pages}{53--63}.
\newblock


\bibitem[\protect\citeauthoryear{Wang and Lo}{Wang and Lo}{2016}]%
        {versionhistoryjsep}
\bibfield{author}{\bibinfo{person}{S. Wang} {and} \bibinfo{person}{D. Lo}.}
  \bibinfo{year}{2016}\natexlab{}.
\newblock \showarticletitle{AmaLgam+: Composing Rich Information Sources for
  Accurate Bug Localization}.
\newblock \bibinfo{journal}{\emph{JSEP}} \bibinfo{volume}{28},
  \bibinfo{number}{10} (\bibinfo{year}{2016}), \bibinfo{pages}{921--942}.
\newblock


\bibitem[\protect\citeauthoryear{Wen, Wu, and Cheung}{Wen
  et~al\mbox{.}}{2016}]%
        {locus}
\bibfield{author}{\bibinfo{person}{M. Wen}, \bibinfo{person}{R. Wu}, {and}
  \bibinfo{person}{S.~C. Cheung}.} \bibinfo{year}{2016}\natexlab{}.
\newblock \showarticletitle{Locus: Locating bugs from software changes}. In
  \bibinfo{booktitle}{\emph{Proc. ASE}}. \bibinfo{pages}{262--273}.
\newblock


\bibitem[\protect\citeauthoryear{Wong, Xiong, Zhang, Hao, Zhang, and Mei}{Wong
  et~al\mbox{.}}{2014}]%
        {brtracer}
\bibfield{author}{\bibinfo{person}{C.~P. Wong}, \bibinfo{person}{Y. Xiong},
  \bibinfo{person}{H. Zhang}, \bibinfo{person}{D. Hao}, \bibinfo{person}{L.
  Zhang}, {and} \bibinfo{person}{H. Mei}.} \bibinfo{year}{2014}\natexlab{}.
\newblock \showarticletitle{Boosting Bug-Report-Oriented Fault Localization
  with Segmentation and Stack-Trace Analysis}. In
  \bibinfo{booktitle}{\emph{Proc. ICSME}}. \bibinfo{pages}{181--190}.
\newblock


\bibitem[\protect\citeauthoryear{Xia, Bao, Lo, and Li}{Xia
  et~al\mbox{.}}{2016}]%
        {harmful}
\bibfield{author}{\bibinfo{person}{X. Xia}, \bibinfo{person}{L. Bao},
  \bibinfo{person}{D. Lo}, {and} \bibinfo{person}{S. Li}.}
  \bibinfo{year}{2016}\natexlab{}.
\newblock \showarticletitle{``Automated Debugging Considered Harmful"
  Considered Harmful: A User Study Revisiting the Usefulness of Spectra-Based
  Fault Localization Techniques with Professionals Using Real Bugs from Large
  Systems}. In \bibinfo{booktitle}{\emph{Proc. ICSME}}.
  \bibinfo{pages}{267--278}.
\newblock


\bibitem[\protect\citeauthoryear{Ye, Bunescu, and Liu}{Ye
  et~al\mbox{.}}{2014}]%
        {learn2rank}
\bibfield{author}{\bibinfo{person}{Xin Ye}, \bibinfo{person}{Razvan Bunescu},
  {and} \bibinfo{person}{Chang Liu}.} \bibinfo{year}{2014}\natexlab{}.
\newblock \showarticletitle{Learning to Rank Relevant Files for Bug Reports
  Using Domain Knowledge}. In \bibinfo{booktitle}{\emph{Proc. FSE}}.
  \bibinfo{pages}{689--699}.
\newblock


\bibitem[\protect\citeauthoryear{Ye, Shen, Ma, Bunescu, and Liu}{Ye
  et~al\mbox{.}}{2016}]%
        {wembedding}
\bibfield{author}{\bibinfo{person}{X. Ye}, \bibinfo{person}{H. Shen},
  \bibinfo{person}{X. Ma}, \bibinfo{person}{R. Bunescu}, {and}
  \bibinfo{person}{C. Liu}.} \bibinfo{year}{2016}\natexlab{}.
\newblock \showarticletitle{From Word Embeddings to Document Similarities for
  Improved Information Retrieval in Software Engineering}. In
  \bibinfo{booktitle}{\emph{Proc. ICSE}}. \bibinfo{pages}{404--415}.
\newblock


\bibitem[\protect\citeauthoryear{Youm, Ahn, Kim, and Lee}{Youm
  et~al\mbox{.}}{2015}]%
        {blia}
\bibfield{author}{\bibinfo{person}{K.~C. Youm}, \bibinfo{person}{J. Ahn},
  \bibinfo{person}{J. Kim}, {and} \bibinfo{person}{E. Lee}.}
  \bibinfo{year}{2015}\natexlab{}.
\newblock \showarticletitle{Bug Localization Based on Code Change Histories and
  Bug Reports}. In \bibinfo{booktitle}{\emph{Proc. APSEC}}.
  \bibinfo{pages}{190--197}.
\newblock


\bibitem[\protect\citeauthoryear{Yuan, Lo, and Lawall}{Yuan
  et~al\mbox{.}}{2014}]%
        {wordsim}
\bibfield{author}{\bibinfo{person}{T. Yuan}, \bibinfo{person}{D. Lo}, {and}
  \bibinfo{person}{J. Lawall}.} \bibinfo{year}{2014}\natexlab{}.
\newblock \showarticletitle{Automated {C}onstruction of a {S}oftware-Specific
  {W}ord {S}imilarity {D}atabase}. In \bibinfo{booktitle}{\emph{Proc.
  CSMR-WCRE}}. \bibinfo{pages}{44--53}.
\newblock


\bibitem[\protect\citeauthoryear{Zhou, Zhang, and Lo}{Zhou
  et~al\mbox{.}}{2012}]%
        {buglocator}
\bibfield{author}{\bibinfo{person}{J. Zhou}, \bibinfo{person}{H. Zhang}, {and}
  \bibinfo{person}{D. Lo}.} \bibinfo{year}{2012}\natexlab{}.
\newblock \showarticletitle{Where should the bugs be fixed? More accurate
  information retrieval-based bug localization based on bug reports}. In
  \bibinfo{booktitle}{\emph{Proc. ICSE}}. \bibinfo{pages}{14--24}.
\newblock


\bibitem[\protect\citeauthoryear{Zhou, Tong, Gu, and Gall}{Zhou
  et~al\mbox{.}}{2014}]%
        {brcls}
\bibfield{author}{\bibinfo{person}{Y. Zhou}, \bibinfo{person}{Y. Tong},
  \bibinfo{person}{R. Gu}, {and} \bibinfo{person}{H. Gall}.}
  \bibinfo{year}{2014}\natexlab{}.
\newblock \showarticletitle{Combining Text Mining and Data Mining for Bug
  Report Classification}. In \bibinfo{booktitle}{\emph{Proc. ICSME}}.
  \bibinfo{pages}{311--320}.
\newblock


\bibitem[\protect\citeauthoryear{Zimmermann, Nagappan, and Zeller}{Zimmermann
  et~al\mbox{.}}{2008}]%
        {zimmer}
\bibfield{author}{\bibinfo{person}{T. Zimmermann}, \bibinfo{person}{N.
  Nagappan}, {and} \bibinfo{person}{A. Zeller}.}
  \bibinfo{year}{2008}\natexlab{}.
\newblock \showarticletitle{Predicting {B}ugs from {H}istory}.
\newblock In \bibinfo{booktitle}{\emph{Software Evolution}}.
  \bibinfo{publisher}{Springer}, \bibinfo{pages}{69--88}.
\newblock


\end{thebibliography}
}
%
%
\end{document}